\documentclass[twocolumn]{aastex631}
\shortauthors{Mali \& Essick}

\usepackage{multirow}
\usepackage{booktabs} 
\usepackage{amsmath}

\newcommand{\PDB}{\textsc{PDB}}
\newcommand{\PDBP}{\textsc{PDB}$\times$\textsc{P}}
\newcommand{\PDBpP}{\textsc{PDB}$+$\textsc{P}}
\newcommand{\MultiPDB}{\textsc{MultiPDB}}
\newcommand{\DoubleDip}{\textsc{DoubleDip}}

\newcommand{\peakone}{\ensuremath{\mu^{\mathrm{peak}}_1}}
\newcommand{\peaktwo}{\ensuremath{\mu^{\mathrm{peak}}_2}}
\newcommand{\peakhatone}{\ensuremath{\hat{\mu}^{7:11}_1}}
\newcommand{\peakhattwo}{\ensuremath{\hat{\mu}^{25:40}_2}}

\newcommand{\sigone}{\ensuremath{\sigma^{\mathrm{peak}}_1}}
\newcommand{\sigtwo}{\ensuremath{\sigma^{\mathrm{peak}}_2}}
\newcommand{\sighatone}{\ensuremath{\hat{\sigma}^{7:11}_1}}
\newcommand{\sighattwo}{\ensuremath{\hat{\sigma}^{25:40}_2}}

\newcommand{\mmax}{\ensuremath{m_{\mathrm{max}}}}
\newcommand{\mninenine}{\ensuremath{m_{99}}}

\newcommand{\gammahightwo}{\ensuremath{\gamma^{\mathrm{high}}_2}}
\newcommand{\gammalowtwo}{\ensuremath{\gamma^{\mathrm{low}}_2}}
\newcommand{\gammahighone}{\ensuremath{\gamma^{\mathrm{high}}_1}}
\newcommand{\gammalowone}{\ensuremath{\gamma^{\mathrm{low}}_1}}

\newcommand{\etahightwo}{\ensuremath{\eta^{\mathrm{high}}_2}}
\newcommand{\etalowtwo}{\ensuremath{\eta^{\mathrm{low}}_2}}
\newcommand{\etahighone}{\ensuremath{\eta^{\mathrm{high}}_1}}
\newcommand{\etalowone}{\ensuremath{\eta^{\mathrm{low}}_1}}
\newcommand{\etamax}{\ensuremath{\eta_{\mathrm{max}}}}
\newcommand{\rPDBmupeakone}{\ensuremath{-}}
\newcommand{\rPDBmuhatone}{\ensuremath{-0.16}}
\newcommand{\rPDBsigmahatone}{\ensuremath{0.17}}
\newcommand{\rPDBmupeaktwo}{\ensuremath{-}}
\newcommand{\rPDBmupeaktwoconditioned}{\ensuremath{-}}
\newcommand{\rPDBmuhattwo}{\ensuremath{-0.38}}
\newcommand{\rPDBsigmahattwo}{\ensuremath{-0.46}}
\newcommand{\rPDBmninenine}{\ensuremath{-0.33}}
\newcommand{\rPDBmmax}{\ensuremath{-0.38}}
\newcommand{\rPDBPmupeakone}{\ensuremath{-}}
\newcommand{\rPDBPmuhatone}{\ensuremath{-0.24}}
\newcommand{\rPDBPsigmahatone}{\ensuremath{0.14}}
\newcommand{\rPDBPmupeaktwo}{\ensuremath{-0.48}}
\newcommand{\rPDBPmupeaktwoconditioned}{\ensuremath{-0.73}}
\newcommand{\rPDBPmuhattwo}{\ensuremath{-0.68}}
\newcommand{\rPDBPsigmahattwo}{\ensuremath{-0.46}}
\newcommand{\rPDBPmninenine}{\ensuremath{-0.29}}
\newcommand{\rPDBPmmax}{\ensuremath{-0.04}}
\newcommand{\rDoubleDipmupeakone}{\ensuremath{-}}
\newcommand{\rDoubleDipmuhatone}{\ensuremath{-0.48}}
\newcommand{\rDoubleDipsigmahatone}{\ensuremath{-0.37}}
\newcommand{\rDoubleDipmupeaktwo}{\ensuremath{-}}
\newcommand{\rDoubleDipmupeaktwoconditioned}{\ensuremath{-}}
\newcommand{\rDoubleDipmuhattwo}{\ensuremath{-0.54}}
\newcommand{\rDoubleDipsigmahattwo}{\ensuremath{-0.09}}
\newcommand{\rDoubleDipmninenine}{\ensuremath{-0.26}}
\newcommand{\rDoubleDipmmax}{\ensuremath{-0.15}}
\newcommand{\rMultiPDBmupeakone}{\ensuremath{-0.49}}
\newcommand{\rMultiPDBmuhatone}{\ensuremath{-0.65}}
\newcommand{\rMultiPDBsigmahatone}{\ensuremath{-0.17}}
\newcommand{\rMultiPDBmupeaktwo}{\ensuremath{-0.4}}
\newcommand{\rMultiPDBmupeaktwoconditioned}{\ensuremath{-0.66}}
\newcommand{\rMultiPDBmuhattwo}{\ensuremath{-0.72}}
\newcommand{\rMultiPDBsigmahattwo}{\ensuremath{-0.37}}
\newcommand{\rMultiPDBmninenine}{\ensuremath{-0.23}}
\newcommand{\rMultiPDBmmax}{\ensuremath{-0.03}}
\newcommand{\rPDBgammahighone}{\ensuremath{-0.19}}
\newcommand{\rPDBPgammahighone}{\ensuremath{-0.2}}
\newcommand{\rDoubleDipgammahighone}{\ensuremath{-0.06}}
\newcommand{\rMultiPDBgammahighone}{\ensuremath{0.08}}
\newcommand{\MultiPDBA}{\ensuremath{0.64^{+0.33}_{-0.56}}}
\newcommand{\MultiPDBHzero}{\ensuremath{64.96^{+42.92}_{-22.67}}}
\newcommand{\MultiPDBOmzero}{\ensuremath{0.41^{+0.53}_{-0.37}}}
\newcommand{\MultiPDBalphaone}{\ensuremath{-4.06^{+1.65}_{-0.83}}}
\newcommand{\MultiPDBalphatwo}{\ensuremath{-1.19^{+0.86}_{-0.59}}}
\newcommand{\MultiPDBbetaone}{\ensuremath{0.59^{+0.85}_{-0.50}}}
\newcommand{\MultiPDBbetatwo}{\ensuremath{2.70^{+1.71}_{-1.11}}}
\newcommand{\MultiPDBetahigh}{\ensuremath{25.98^{+21.70}_{-22.86}}}
\newcommand{\MultiPDBetalow}{\ensuremath{26.42^{+21.24}_{-22.65}}}
\newcommand{\MultiPDBetamax}{\ensuremath{4.62^{+4.67}_{-3.83}}}
\newcommand{\MultiPDBetamin}{\ensuremath{37.47^{+11.24}_{-11.25}}}
\newcommand{\MultiPDBgammahigh}{\ensuremath{6.41^{+1.49}_{-2.13}}}
\newcommand{\MultiPDBgammalow}{\ensuremath{3.01^{+0.87}_{-0.62}}}
\newcommand{\MultiPDBmbreak}{\ensuremath{3.97^{+0.92}_{-1.43}}}
\newcommand{\MultiPDBmmax}{\ensuremath{72.01^{+25.05}_{-33.45}}}
\newcommand{\MultiPDBmmin}{\ensuremath{0.86^{+0.31}_{-0.33}}}
\newcommand{\MultiPDBmupeakone}{\ensuremath{9.10^{+0.83}_{-1.56}}}
\newcommand{\MultiPDBmupeaktwo}{\ensuremath{31.94^{+7.64}_{-5.43}}}
\newcommand{\MultiPDBpeakconstantone}{\ensuremath{49.20^{+42.53}_{-30.85}}}
\newcommand{\MultiPDBpeakconstanttwo}{\ensuremath{62.19^{+33.64}_{-41.73}}}
\newcommand{\MultiPDBsigpeakone}{\ensuremath{1.33^{+1.38}_{-0.30}}}
\newcommand{\MultiPDBsigpeaktwo}{\ensuremath{6.51^{+18.80}_{-3.87}}}
\newcommand{\MultiPDBmuhatvalsonezero}{\ensuremath{8.95^{+0.41}_{-0.38}}}
\newcommand{\MultiPDBsigmahatvalsonezero}{\ensuremath{0.98^{+0.11}_{-0.10}}}
\newcommand{\MultiPDBmuhatvals}{\ensuremath{31.64^{+1.19}_{-1.60}}}
\newcommand{\MultiPDBsigmahatvals}{\ensuremath{4.01^{+0.23}_{-0.78}}}
\newcommand{\MultiPDBmninenine}{\ensuremath{44.90^{+15.76}_{-10.31}}}
\newcommand{\DoubleDipA}{\ensuremath{0.92^{+0.08}_{-0.37}}}
\newcommand{\DoubleDipAtwo}{\ensuremath{0.86^{+0.11}_{-0.29}}}
\newcommand{\DoubleDipHzero}{\ensuremath{71.44^{+38.58}_{-28.21}}}
\newcommand{\DoubleDipOmzero}{\ensuremath{0.41^{+0.52}_{-0.38}}}
\newcommand{\DoubleDipalphaone}{\ensuremath{-1.75^{+2.68}_{-2.61}}}
\newcommand{\DoubleDipalphatwo}{\ensuremath{-1.55^{+1.85}_{-0.87}}}
\newcommand{\DoubleDipbetaone}{\ensuremath{0.89^{+0.94}_{-0.67}}}
\newcommand{\DoubleDipbetatwo}{\ensuremath{2.69^{+1.85}_{-1.16}}}
\newcommand{\DoubleDipetahigh}{\ensuremath{30.80^{+17.21}_{-21.42}}}
\newcommand{\DoubleDipetahightwo}{\ensuremath{28.78^{+19.26}_{-23.83}}}
\newcommand{\DoubleDipetalow}{\ensuremath{28.15^{+19.77}_{-20.95}}}
\newcommand{\DoubleDipetalowtwo}{\ensuremath{32.83^{+15.57}_{-21.53}}}
\newcommand{\DoubleDipetamax}{\ensuremath{5.90^{+3.88}_{-3.32}}}
\newcommand{\DoubleDipetamin}{\ensuremath{37.48^{+11.09}_{-11.13}}}
\newcommand{\DoubleDipgammahigh}{\ensuremath{7.24^{+0.70}_{-1.94}}}
\newcommand{\DoubleDipgammahightwo}{\ensuremath{28.13^{+31.32}_{-8.41}}}
\newcommand{\DoubleDipgammalow}{\ensuremath{2.61^{+0.88}_{-0.28}}}
\newcommand{\DoubleDipgammalowtwo}{\ensuremath{10.98^{+2.67}_{-1.44}}}
\newcommand{\DoubleDipmbreak}{\ensuremath{3.39^{+1.42}_{-1.23}}}
\newcommand{\DoubleDipmmax}{\ensuremath{40.68^{+29.91}_{-5.31}}}
\newcommand{\DoubleDipmmin}{\ensuremath{0.86^{+0.31}_{-0.32}}}
\newcommand{\DoubleDipmuhatvalsonezero}{\ensuremath{8.83^{+0.29}_{-0.29}}}
\newcommand{\DoubleDipsigmahatvalsonezero}{\ensuremath{1.04^{+0.10}_{-0.20}}}
\newcommand{\DoubleDipmuhatvals}{\ensuremath{31.33^{+1.06}_{-0.49}}}
\newcommand{\DoubleDipsigmahatvals}{\ensuremath{4.13^{+0.11}_{-0.19}}}
\newcommand{\DoubleDipmninenine}{\ensuremath{45.59^{+13.08}_{-9.51}}}
\newcommand{\PDBPA}{\ensuremath{0.92^{+0.07}_{-0.25}}}
\newcommand{\PDBPHzero}{\ensuremath{65.60^{+44.69}_{-23.12}}}
\newcommand{\PDBPOmzero}{\ensuremath{0.43^{+0.50}_{-0.39}}}
\newcommand{\PDBPalphaone}{\ensuremath{-1.21^{+2.14}_{-1.61}}}
\newcommand{\PDBPalphatwo}{\ensuremath{-2.29^{+0.45}_{-0.47}}}
\newcommand{\PDBPbetaone}{\ensuremath{0.96^{+0.99}_{-0.70}}}
\newcommand{\PDBPbetatwo}{\ensuremath{2.51^{+1.76}_{-1.16}}}
\newcommand{\PDBPetahigh}{\ensuremath{30.59^{+17.42}_{-20.23}}}
\newcommand{\PDBPetalow}{\ensuremath{27.77^{+19.90}_{-19.85}}}
\newcommand{\PDBPetamax}{\ensuremath{3.50^{+5.84}_{-3.22}}}
\newcommand{\PDBPetamin}{\ensuremath{37.57^{+11.24}_{-11.39}}}
\newcommand{\PDBPgammahigh}{\ensuremath{6.38^{+1.24}_{-1.23}}}
\newcommand{\PDBPgammalow}{\ensuremath{2.56^{+0.60}_{-0.24}}}
\newcommand{\PDBPmbreak}{\ensuremath{3.60^{+1.27}_{-1.47}}}
\newcommand{\PDBPmmax}{\ensuremath{79.39^{+18.74}_{-36.00}}}
\newcommand{\PDBPmmin}{\ensuremath{0.86^{+0.31}_{-0.32}}}
\newcommand{\PDBPmupeak}{\ensuremath{34.74^{+8.03}_{-5.44}}}
\newcommand{\PDBPpeakconstant}{\ensuremath{32.25^{+47.62}_{-19.12}}}
\newcommand{\PDBPsigpeak}{\ensuremath{4.25^{+16.18}_{-2.82}}}
\newcommand{\PDBPmuhatvalsonezero}{\ensuremath{8.67^{+0.26}_{-0.08}}}
\newcommand{\PDBPsigmahatvalsonezero}{\ensuremath{1.12^{+0.01}_{-0.08}}}
\newcommand{\PDBPmuhatvals}{\ensuremath{31.90^{+1.69}_{-1.67}}}
\newcommand{\PDBPsigmahatvals}{\ensuremath{3.98^{+0.34}_{-1.01}}}
\newcommand{\PDBPmninenine}{\ensuremath{42.52^{+12.98}_{-8.23}}}
\newcommand{\PDBA}{\ensuremath{0.88^{+0.11}_{-0.45}}}
\newcommand{\PDBHzero}{\ensuremath{68.71^{+43.93}_{-26.20}}}
\newcommand{\PDBOmzero}{\ensuremath{0.46^{+0.48}_{-0.42}}}
\newcommand{\PDBalphaone}{\ensuremath{-2.11^{+2.03}_{-1.95}}}
\newcommand{\PDBalphatwo}{\ensuremath{-1.75^{+0.35}_{-0.30}}}
\newcommand{\PDBbetaone}{\ensuremath{0.96^{+1.01}_{-0.68}}}
\newcommand{\PDBbetatwo}{\ensuremath{2.73^{+2.01}_{-1.27}}}
\newcommand{\PDBetahigh}{\ensuremath{30.06^{+17.82}_{-21.01}}}
\newcommand{\PDBetalow}{\ensuremath{27.30^{+20.25}_{-21.06}}}
\newcommand{\PDBetamax}{\ensuremath{5.53^{+3.71}_{-3.01}}}
\newcommand{\PDBetamin}{\ensuremath{37.46^{+11.33}_{-11.28}}}
\newcommand{\PDBgammahigh}{\ensuremath{6.05^{+1.38}_{-1.37}}}
\newcommand{\PDBgammalow}{\ensuremath{2.66^{+0.88}_{-0.33}}}
\newcommand{\PDBmbreak}{\ensuremath{3.05^{+1.71}_{-0.97}}}
\newcommand{\PDBmmax}{\ensuremath{58.42^{+25.05}_{-15.58}}}
\newcommand{\PDBmmin}{\ensuremath{0.87^{+0.30}_{-0.33}}}
\newcommand{\PDBmuhatvalsonezero}{\ensuremath{8.74^{+0.16}_{-0.05}}}
\newcommand{\PDBsigmahatvalsonezero}{\ensuremath{1.13^{+0.01}_{-0.05}}}
\newcommand{\PDBmuhatvals}{\ensuremath{31.35^{+0.17}_{-0.23}}}
\newcommand{\PDBsigmahatvals}{\ensuremath{4.24^{+0.03}_{-0.06}}}
\newcommand{\PDBmninenine}{\ensuremath{46.88^{+13.58}_{-11.10}}}
\newcommand{\MultiPDBpeakconstantoneNORMED}{\ensuremath{13.55^{+14.75}_{-8.80}}}
\newcommand{\MultiPDBpeakconstanttwoNORMED}{\ensuremath{3.76^{+3.89}_{-3.04}}}
\newcommand{\PDBPpeakconstantNORMED}{\ensuremath{3.10^{+4.83}_{-2.25}}}
\newcommand{\statPDBr}{\ensuremath{-0.009}}
\newcommand{\statPDBmathcalR}{\ensuremath{0.87}}
\newcommand{\statPDBPr}{\ensuremath{0.146}}
\newcommand{\statPDBPmathcalR}{\ensuremath{7.92}}
\newcommand{\statDoubleDipr}{\ensuremath{0.242}}
\newcommand{\statDoubleDipmathcalR}{\ensuremath{19.17}}
\newcommand{\statMultiPDBr}{\ensuremath{0.467}}
\newcommand{\statMultiPDBmathcalR}{\ensuremath{126.28}}
\newcommand{\rDoubleDipgammahightwo}{\ensuremath{0.18}}
\newcommand{\rDoubleDipgammalowtwo}{\ensuremath{-0.13}}
\newcommand{\rDoubleDipetamax}{\ensuremath{0.27}}
\newcommand{\cutpeak}{\ensuremath{8}}

\newcommand{\numevents}{63}
\newcommand{\peakatten}{9$\rm M_{\odot}$}
\newcommand{\peakatthirty}{32$\rm M_{\odot}$}
\newcommand{\shelf}{46$\rm M_{\odot}$}
\newcommand{\lowerMGlow}{3.0$\rm M_{\odot}$}
\newcommand{\lowerMGhigh}{6.4$\rm M_{\odot}$}
\begin{document}
\title{
Striking a Chord with Spectral Sirens: \\ multiple features in the compact binary population correlate with $H_0$
}

\author{Utkarsh Mali}
\email{utkarsh.mali@utoronto.ca}
\affiliation{Canadian Institute for Theoretical Astrophysics, 60 St. George St, Toronto, ON M5S 3H8, Canada}
\affiliation{Department of Physics, University of Toronto, 60 St. George St, Toronto, ON M5S 1A7, Canada}

\author{Reed Essick}
\affiliation{Canadian Institute for Theoretical Astrophysics, 60 St. George St, Toronto, ON M5S 3H8, Canada}
\affiliation{Department of Physics, University of Toronto, 60 St. George St, Toronto, ON M5S 1A7, Canada}
\affiliation{David A. Dunlap Department of Astronomy, University of Toronto, 50 St. George Street, Toronto, ON M5S 3H4, Canada}

\begin{abstract}
    Spectral siren measurements of the Hubble constant ($H_0$) rely on correlations between observed detector-frame masses and luminosity distances. Features in the source-frame mass distribution can induce these correlations.
    It is crucial, then, to understand (i) which features in the source-frame mass distribution are robust against model (re)parametrization, (ii) which features carry the most information about $H_0$, and (iii) whether distinct features independently correlate with cosmological parameters.
    We study these questions using real gravitational-wave observations from the LIGO-Virgo-KAGRA Collaborations' third observing run.
    Although constraints on $H_0$ are weak, we find that current data reveals several prominent features in the mass distribution, including peaks in the binary black hole source-frame mass distribution near $\sim$ \peakatten~and $\sim$ \peakatthirty~and a roll-off at masses above $\sim$ \shelf.
    For the first time using real data, we show that all of these features carry cosmological information and that the peak near $\sim$ \peakatthirty~consistently correlates with $H_0$ most strongly.
    Introducing model-independent summary statistics, we show that these statistics independently correlate with $H_0$, exactly what is required to limit systematics within future spectral siren measurements from the (expected) astrophysical evolution of the mass distribution.
\end{abstract}


\section{Introduction}
\label{sec:intro}

The observation of gravitational waves (GWs) from compact binary coalescences (CBCs) observed with the advanced LIGO, Virgo, and KAGRA (LVK) detectors \citep{Acernese_2015, Aasi_2015,kagra} provide a new window onto many astrophysical and cosmological phenomena \citep{Abbott_2016, Abbott_20162, Abbott_2019, theligoscientificcollaboration2021gwtc3, Abbott_2021, theligoscientificcollaboration2022gwtc21deepextendedcatalog, Abbott_2023}.
Many authors have proposed various methods to use CBCs as tracers of the Hubble relation. While the specific approaches vary \citep{abbott2017gw170817, Soares-Santos_2019, Ezquiaga_2022}, they all involve obtaining simultaneous estimates of both the luminosity distance ($D_L$) and redshift ($z$) for a set of sources.
These estimates are then used to fit the Hubble relation $H(z)$, allowing us to extract key cosmological parameters.

Several electromagnetic (EM) approaches operate in this way; these include Hubble's original observations \citep{hubble1929relation}, recent catalogues of cepheids \citep{Riess_2020}, the tip of the red-giant branch \citep{lee1993tip, freedman2020calibration} and the J-region asymptotic giant branch \citep{freedman2023progress}.
Often, with EM measurements, it is possible to precisely measure $z$, but more difficult to estimate $D_L$.
Indeed, much of the discussion in the literature focuses on different ways to calibrate the local distance ladder and thereby improve estimates of $D_L$ \citep{freedman2024statusreportchicagocarnegiehubble}.

Conversely, GW observations of CBCs can directly constrain $D_L$ independently of the distance ladder and without relying on other observations~\citep{schutz1986determining}.
However, it is much more difficult to reliably estimate z for GW sources.
In General Relativity (GR), vacuum solutions to the Einstein field equations do not automatically contain a fixed scale  (like the known rest-frame frequencies of atomic and molecular lines in EM observations) that can be used to measure $z$.
As such, it is relatively easy to measure $D_L$ with GW observations, and many approaches in GW cosmology focus on different ways of obtaining estimates of $z$.

Several of the most common approaches rely on EM data to obtain $z$.
\emph{Bright sirens} use observations of EM counterparts for individual events to measure $z$ through an association with a host galaxy \citep{Holz_2005, PhysRevD.74.063006, Nissanke_2010, nissanke2013determininghubbleconstantgravitational, abbott2017gw170817}.
\emph{Dark sirens} instead do not rely on the identification of an EM counterpart for individual events but rather probabilistically associate CBCs with catalogues of potential host galaxies \citep{PhysRevD.86.043011, Fishbach_2019_ds, soares2019first, PhysRevD.101.122001, abbott2021gravitational, 10.1093/mnras/stac366, Mukherjee_2021, PhysRevD.108.042002, mastrogiovanni2023novelapproachinferpopulation, gair2023hitchhiker, hanselman2024gravitationalwavedarksirencosmology,mukherjee2024crosscorrelatingdarksirensgalaxies}.
In both cases, EM observations effectively serve as a separate source of information about $z$ for individual events.

However, by considering a catalogue of CBCs, information about $z$ can be inferred from GW data alone.
That is, GW cosmology does not need to rely on EM data. 
Several approaches have been proposed (see, e.g., \cite{Chatterjee_2021, Ezquiaga_2022, li2024multispectralsirensgravitationalwavecosmology}).
In general, they rely on identifying an observable feature of individual CBC systems that correlates with the source-frame mass ($m_s$).
Measurement of that parameter then provides information about $m_s$, and the GW data directly constrains the detector-frame (or redshifted) mass: $m_d = (1+z) m_s$.
Joint constraints on $m_d$ and $m_s$ thereby provide an estimate for $z$.

We focus on \emph{spectral sirens}, in which features in the distribution of source-frame masses inferred from a catalogue of CBCs provide additional information about $m_s$ for individual events.
This approach has been studied in several contexts, including for binary neutron star (BNS) and binary black hole (BBH) systems with current and proposed GW detectors \citep{schutz1986determining, Chernoff_1993, Messenger:2011gi, taylor2012cosmology, Chen_2018, Farr_2019, PhysRevD.101.122001, Chatterjee_2021, You_2021, Mastrogiovanni_2021, Mastrogiovanni_2022, Mancarella_2022, Karathanasis:2022rtr, Leyde_2022, Ezquiaga_2022, PhysRevD.108.042002, mastrogiovanni2023novelapproachinferpopulation, pierra2024study, hernandez2024gapsbumpsspectralsiren, farah2024needknowastrophysicsfreegravitationalwave}.

Typically, authors assume the presence of a feature in the source-frame mass distribution, like a narrow peak \citep{taylor2012cosmology} or a rapid fall-off \citep{Farr_2019, 10.3389/fspas.2020.00038}, which then induces a tell-tale pattern in the joint distribution of detector-frame masses and luminosity distance.
See, e.g., \citet{Ezquiaga_2022} for a review.

The efficacy of spectral sirens has been simulated by many authors \citep{taylor2012cosmology, Ezquiaga_2022, farah2024needknowastrophysicsfreegravitationalwave, ray2024searchingbinaryblackhole}, but several questions remain.
While the method may work in simulated universes, we wish to study its prospects with real observations.
As such, we study
\begin{itemize}
    \item whether there are useful features in the source-frame mass distribution consistently inferred with different mass models,
    \item which of these features (if any) carry the most information about $H(z)$,
    \item whether multiple features correlate equally and independently with $H(z)$ or if most of the information is associated with a single feature.
\end{itemize}
Since a potential weakness of spectral sirens is the unknown astrophysical evolution of the source-frame mass with time, the last point is of particular interest.
That is, if the distribution of $m_s$ is not independent of $z$, then correlations between $m_d$ and $z$ may not be due to the universe's expansion, but instead simply associated with stellar astrophysics (e.g., changes in binary evolution).
However, if the source-frame mass distribution contains multiple features across mass scales, it is improbable that the same astrophysical processes would affect all of them identically, whereas cosmological effects would.
Put differently, stellar evolution operates differently at different mass scales but cosmology redshifts all GWs in the same way. 
Therefore, if multiple features in the source-frame mass distribution separately correlate with $H_0$, then it is likely we will be able to break the degeneracy between astrophysical evolution of the source-frame mass distribution and $H(z)$.
Our work shows that the source-frame mass distribution inferred from real (observed) GW data has multiple such features.

We begin by outlining our methodology and the observations used in Sec.~\ref{sec:methods}.
We examine the behaviour of several different models of the source-frame mass distribution, each of which is described in Sec.~\ref{sec:mass distributions} (see also Appendix~\ref{sec:pop_models}).
Sec.~\ref{sec:H0 constraints} briefly discusses conclusions from the joint posteriors obtained by simultaneously inferring a $\Lambda$CDM cosmology and our mass distribution before Sec.~\ref{sec:best_feature} examines how different features within the mass distribution encode information about $H_0$.
We identify which parameters correlate most strongly with $H_0$, explore which types of behaviours lead to stronger correlations, and introduce several model-independent summary statistics.
Interestingly, the summary statistics can encode cosmological information even better than individual model parameters.
We show that current data supports the presence of local overdensities (``peaks'') in the source-frame mass distribution near $\sim$ \peakatten~and $\sim$ \peakatthirty~along with a ``roll-off'' near $\sim$ \shelf~and that all of these features correlate with $H_0$.
Sec.~\ref{sec:other_features} then investigates whether these features are independent. We show that their correlations are predominantly driven by their independent correlations with $H_0$ which is essential for a robust spectral siren constraint.
We conclude in Sec.~\ref{sec:discussion}.


\section{Methodology}
\label{sec:methods}

We construct a hierarchical model and use \numevents~confidently detected CBCs from the LVK's third observing run (O3) to simultaneously infer the component mass distribution and cosmological parameters.
We assume CBCs follow an inhomogeneous Poisson process, marginalizing over the overall rate of mergers. The likelihood of observed date ($D_i$) for each event ($i$), given parameters that describe the CBC merger density and the Hubble expansion $H(z)$ (and other population parameters, $\Lambda$) is then 
\begin{equation}\label{eq:likelihood}
    p(\{D_i\}| \Lambda) \propto \frac{1}{\mathcal{E}^{N}}\prod_{i=1}^{N}\mathcal{Z}_{i} 
\end{equation}
where we have defined the single-event evidence $\mathcal{Z}_i$
\begin{equation}\label{eq:Zi}
    \mathcal{Z}_{i}(\Lambda) = p(D_i|\Lambda) = \int p(D_i|\theta) p(\theta|\Lambda)  d\theta
\end{equation}
and the detection probability $\mathcal{E}$
\begin{equation}\label{eq:E}
    \mathcal{E}(\Lambda) = P(\mathrm{det}|\Lambda) = \int P(\mathrm{det}|\theta) p(\theta|\Lambda)  d\theta
\end{equation}
Within these integrals, $\theta$ represents the single-event parameters, like component masses and spins, $p(D_i|\theta)$ is the likelihood of obtaining $D_i$ given a signal described by $\theta$, and $P(\mathrm{det}|\theta)$ is the probability that a signal described by $\theta$ would be detected~\citep[marginalized over noise realizations;][]{essick2023dagnabbitensuringconsistencynoise, essick2023semianalyticsensitivityestimatescatalogs}.
See \cite{skilling2004aip, mandel2010parameter, Thrane_Talbot_2019, mandel2019extracting, Vitale_2021} for reviews.

Our set of \numevents~confident events from O3 was obtained by selecting those events for which at least one search (either cWB \citep{klimenko_2021_4419902}, GstLAL \citep{cannon2020gstlalsoftwareframeworkgravitational}, MBTA \citep{Aubin_2021}, and one of two PyCBC searches \citep{alex_nitz_2024_10473621}) reported a false alarm rate (FAR) $\leq 1/\mathrm{year}$.
We only use events from O3 because real search sensitivity estimates \citep{zenodoO3mcvt} are not available for other observing runs, and O3 contains the vast majority of the publicly available surveyed volume-time (and detected events).

Additionally, we consider events across the entire mass spectrum, including BNS, neutron star-black hole (NSBH), and BBH coalescences.
Previous analyses have focused only on BBHs, applying \textit{ad hoc} cuts based on the secondary mass inferred assuming a reference cosmology.
\textit{De facto}, they assume neutron stars (NSs) are uninformative \textit{a priori}.
Although Sec.~\ref{sec:H0 constraints} finds that much of the information about cosmology from the current catalogue is indeed associated with features at high masses, we also consider low-mass systems. 

For our cosmological model, we assume flat $\Lambda$CDM.
We fit for $H_0$ and the present-day matter density ($\Omega_m$).
The present-day radiation density ($\Omega_r$) is fixed at $0.001$.
We assume the closure condition so that the present-day dark-energy density is $\Omega_\Lambda = 1 - \Omega_m - \Omega_r$. 
Appendix~\ref{sec:priors} lists all priors assumed within our inference.


\begin{figure*}
    \includegraphics[width=0.6\textwidth]{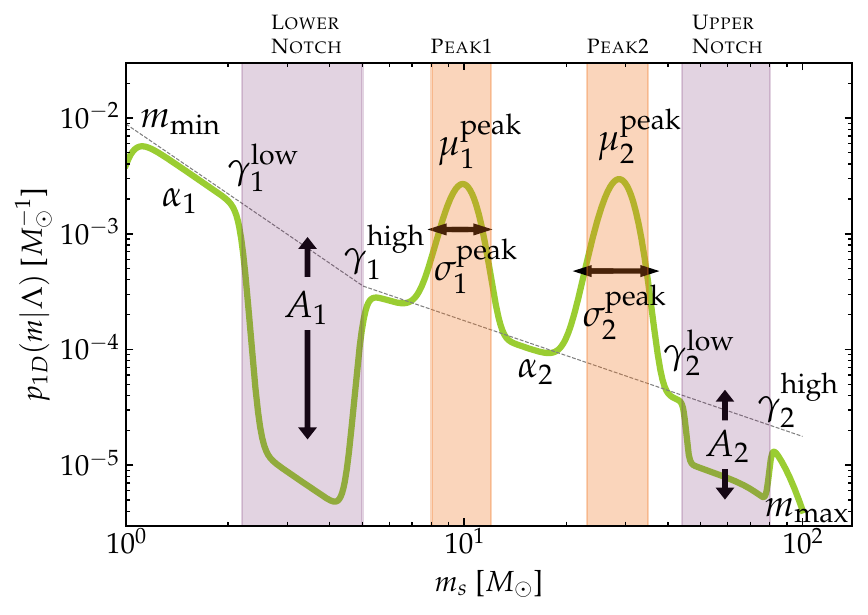}
    \includegraphics[width=0.4\textwidth]{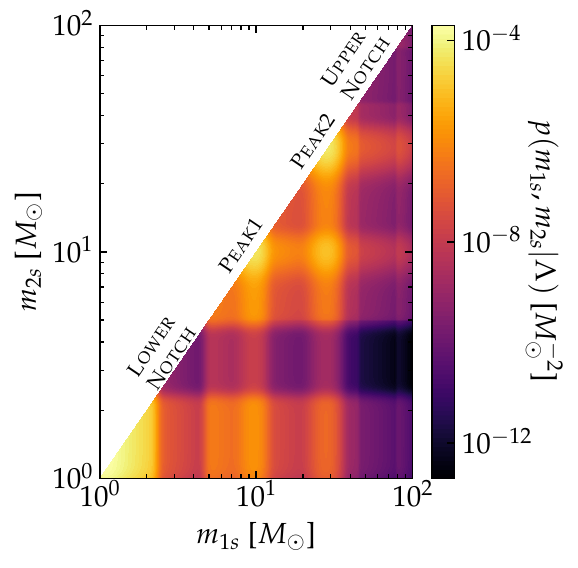}
    \caption{
        (\emph{left}) Schematic representation of $p_{1D}(m|\Lambda)$ (Eq.~\ref{eq:1d-mass}) and associated hyper-parameters.
        The model (\emph{green solid}) is based on a broken power law (\emph{black dotted}) with roll-offs at both high- ($m_\mathrm{max}$) and low-masses ($m_\mathrm{min}$).
        Additional Butterworth notch filters (\emph{purple}) and Gaussian peaks (\emph{orange}) are included as multiplicative factors.
        See Appendix~\ref{sec:pop_models} for details.
        (\emph{right}) The associated joint distribution $p(m_{1s}, m_{2s}|\Lambda)$ (Eq.~\ref{eq:joint mass}).
        Note that features in $p_{1D}$ appear in both $m_{1s}$ and $m_{2s}$ \citep{farah2023kindcomparingbigsmall}.
    }
    \label{fig:params}
\end{figure*}

\subsection{Population Model and Mass Distributions}
\label{sec:mass distributions}

Within our analysis, the distribution of compact binaries is of primary importance.
We focus on the source-frame mass distribution, making simple assumptions for the spin and redshift ($z$) distributions.
We assume fixed distributions over component spins (uniform in magnitude and isotropic in orientation) and that binaries' redshift follows the local star formation rate ($\Phi(z)$, see Appendix~\ref{sec:pop_models}).
\begin{equation}\label{eq:p of z}
    p(z|\Lambda) \propto \frac{dV_c}{dz}\frac{1}{1+z} \Phi(z)
\end{equation}
Here, $V_c$ is the comoving volume and the factor, and $(1+z)^{-1}$ accounts for cosmological time-dilation between the source and detector frames.
Note that $p(z|\Lambda)$ implicitly depends on $H(z)$ through $V_c$. 

We extend a phenomenological source-frame mass distribution first introduced in \citet{Fishbach_2020_matter} and later used in \citet{farah2022bridging} and \citet{theligoscientificcollaboration2022populationmergingcompactbinaries}. 
Specifically, we model the joint distribution of source-frame component masses ($m_{1s}$ and $m_{2s}$) as
\begin{multline}\label{eq:joint mass}
    p(m_{1s}, m_{2s}|\Lambda) \propto p_{1D}(m_{1s}|\Lambda) p_{1D}(m_{2s}|\Lambda) \\ \times f(m_{1s}, m_{2s}; \Lambda) \Theta(m_{1s} \geq m_{2s})
\end{multline}
where $p_{1D}$ is a one-dimensional distribution over mass,\footnote{Importantly, $p_{1D}(m|\Lambda)$ should \emph{not} be confused with the marginal distributions over the primary or secondary masses $p(m_1), p(m_2)$.} $f$ is a pairing function that influences how often different component masses form binaries, and $\Theta$ is an indicator function that enforces our labelling convention: $m_{1s} \geq m_{2s}$.
We truncate $p_{1D}$ below 1 $M_{\odot}$ but we do consider high-mass binaries ($m_{1s} > 100\,\rm M_\odot$) to which the LVK is currently sensitive to.
Fig.~\ref{fig:params} shows an example of $p_{1D}(m|\Lambda)$ and the corresponding $p(m_{1s}, m_{2s}|\Lambda)$.

Details of the exact functional forms and priors assumed within the inference are provided in Appendix~\ref{sec:pop_models}, but, briefly, we model $p_{1D}$ as a broken power-law with roll-offs at both high- and low-masses which is further augmented by a set of multiplicative filters that either remove notches or add peaks to the mass distribution.
Specifically, we consider up to two Gaussian peaks and two Butterworth notch filters.
These features give the mass model considerable flexibility while still providing convenient ways to downselect or ``turn off'' specific features.
Secs.~\ref{sec:pdb}-\ref{sec:multiple peaks} describe our models in more detail.

Before performing the integrals in Eqs.~\ref{eq:Zi} and~\ref{eq:E}, we transform our model from source-frame mass and redshift to detector-frame mass ($m_d = m_s (1+z)$) and luminosity distance ($D_L(z)$).
This allows us to conveniently incorporate the cosmological dependence of these transformations within our analysis and approximate our integrals (Eqs.~\ref{eq:Zi} and~\ref{eq:E}) as Monte Carlo samples over parameters that are directly measured ($m_{1d}$, $m_{2d}$, and $D_L$).
That is, we consider
\begin{align}
    p(\theta|\Lambda)
      & = p(m_{1d}, m_{2d}, D_L, \vec{s}_1, \vec{s}_2|\Lambda) \nonumber \\
      & =  p\left.\left(m_{1s}=\frac{m_{1d}}{1+z}, m_{2s}=\frac{m_{2d}}{1+z}\right|\Lambda\right) (1+z)^{-2} \nonumber \\
        & \quad \quad \times p(z=z(D_z)|\Lambda) \left|\frac{dD_L}{dz}\right|^{-1} \nonumber \\
        & \quad \quad \times p(\vec{s}_1, \vec{s}_2|\Lambda)
\end{align}

\begin{figure*}
    \centering
    \includegraphics[width=0.9\textwidth, clip=True, trim=0.0cm 0.25cm 0.0cm 0.75cm]{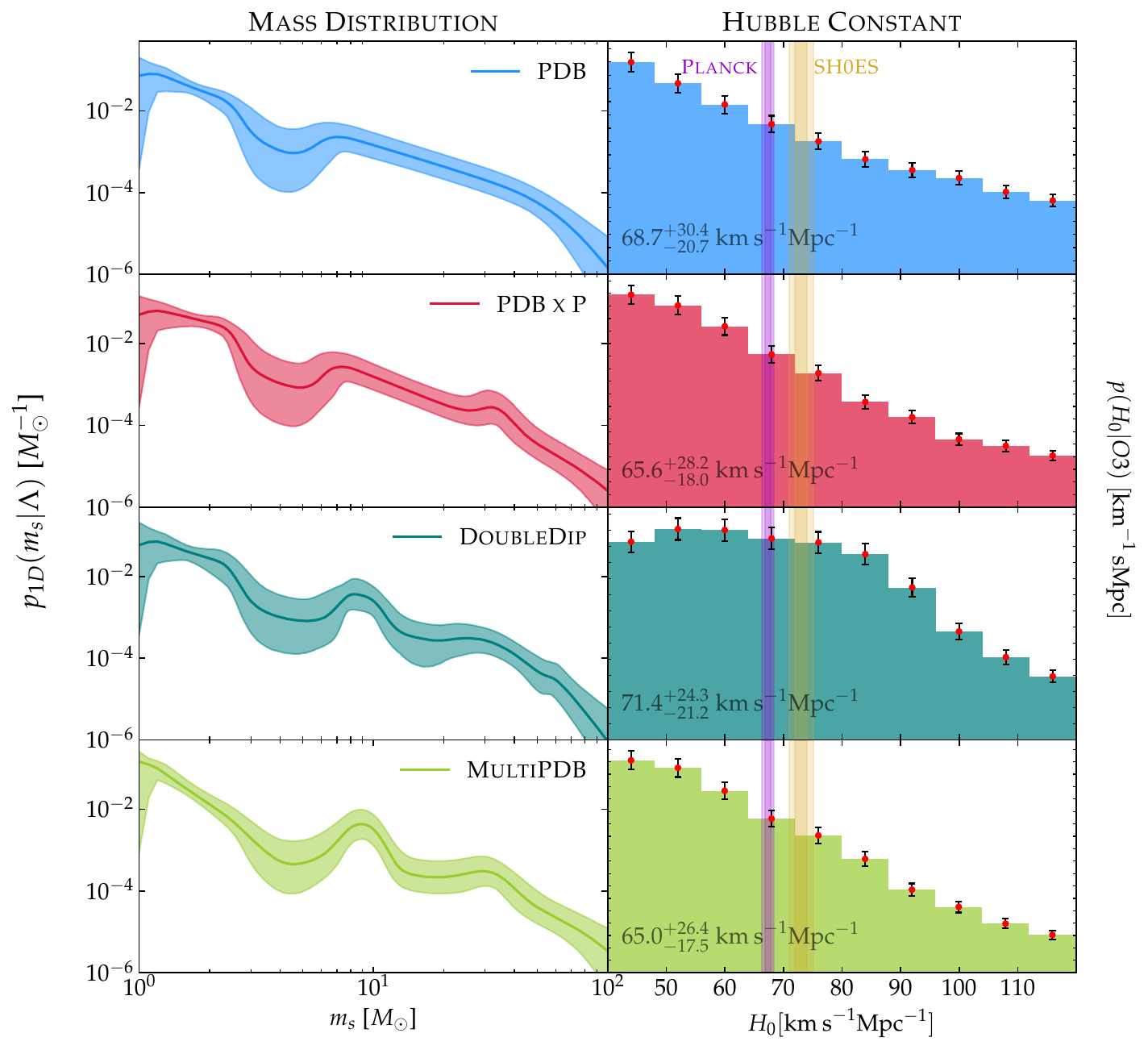}
    \caption{
        (\emph{left}) Posterior medians and 90\% symmetric credible regions for $p_{1D}(m_s|\Lambda)$ as a function of mass conditioned on the O3 catalog for (\emph{top to bottom}) \PDB~(Sec.~\ref{sec:pdb}), \PDBP~(Sec.~\ref{sec:pdbp}), \DoubleDip~and \MultiPDB~(Sec.~\ref{sec:multiple peaks}).
        In general, $p_{1D}$ decreases as a function of mass, but models that allow for additional flexibility often find evidence for peaks at $\sim$ \peakatten and $\sim$ \peakatthirty.
        (\emph{right}) Corresponding posterior distributions (linear scale) for the Hubble parameter ($H_0$) along with 2-$\sigma$ error bars from the finite number of samples used.
        Median values and 90\% symmetric credible regions are shown in each panel.
        We also show the 1- and 2-$\sigma$ constraints from the Planck and SH0ES collaborations as vertical bars \citep{planck2018, Riess_2022}.
        Although mostly uninformative, the data primarily disfavors large values of $H_0$.
        See Table~\ref{tab:prior} for priors. 
    }
    \label{fig:results}
\end{figure*}

We consider several variations of $p_{1D}(m|\Lambda)$, each with different subsets of features ``switched on.''
In order of increasing complexity, the following sections explain the motivation for each.
Fig.~\ref{fig:results} shows their corresponding posterior distributions conditioned on the LVK's O3 catalogue.

\subsubsection{Base Distribution (\PDB)}
\label{sec:pdb}
First introduced in \citet{Fishbach_2020_matter} and later used in~\citet{farah2022bridging} and~\citet{theligoscientificcollaboration2022populationmergingcompactbinaries}, the Power-law + Dip + Break (\PDB; top row of Fig.~\ref{fig:results}) mass distribution is the simplest $p_{1D}$ model we consider. 
This model includes a broken power-law (indexes $\alpha_1$ and $\alpha_2$ in Fig.~\ref{fig:params}) with high- and low-mass roll-offs ($m_\mathrm{max}$ and $m_\mathrm{min}$) along with a single Butterworth notch (\gammalowone, \gammahighone, and $A_1$) to account for the dearth of compact objects observed between 3-5$\, \rm M_\odot$.
We include this model to make comparisons with previous results and to provide a baseline for how much cosmological information is contained in a source-frame mass distribution that does not have any pronounced peaks.


\subsubsection{Distributions with a Single Peak (\PDBP)}
\label{sec:pdbp}

We also consider a model with a single additional peak (\PDBP: \peaktwo, and \sigtwo~in Fig.~\ref{fig:params}; second row of Fig.~\ref{fig:results}).
There is strong evidence that there is a local over-density in the mass distribution around $30 M_{\odot}$~compared to a single power-law  \citep{theligoscientificcollaboration2022populationmergingcompactbinaries, Callister_2024, farah2023thingsbumpnightassessing}, and \PDBP~models this with a Gaussian peak (see Appendix~\ref{sec:pdb_p}).
We generally find consistent behavior between \PDBP~and other models that include a single peak \citep{theligoscientificcollaboration2022populationmergingcompactbinaries, lvkcosmoO3}

Additionally, while our implementation includes additional features as multiplicative factors, it is common to include such features as additional components in a mixture model \citep{Zevin_2021, Abbott_2019, Abbott_2021, jaxengodfrey2024cosmiccousinsidentificationsubpopulation}.
While these models are not exactly equivalent, both approaches produce similar behaviour.
We demonstrate this in Appendix~\ref{sec:pdb_p} by comparing \PDBP~to a mixture model of \PDB~with a separate Gaussian (\PDBpP).


\subsubsection{Distributions with Multiple Peaks (\DoubleDip~and \MultiPDB)}
\label{sec:multiple peaks}

In addition to an overdensity at $\sim30\,\rm M_\odot$, there is growing evidence in favor of another feature near $\sim10\,\rm M_\odot$ 
\citep{farah2023thingsbumpnightassessing, Callister_2024, ray2024searchingbinaryblackhole, jaxengodfrey2024cosmiccousinsidentificationsubpopulation}.
We introduce two models that try to capture this behaviour.

First, \DoubleDip~(third row in Fig.~\ref{fig:results}) introduces a second Butterworth notch filter (\gammalowtwo, \gammahightwo, and $A_2$ in Fig.~\ref{fig:params}) because two peaks (local maxima) can be achieved by introducing a single notch (local minimum).
Generally, the additional notch produces local maxima in $p_{1D}$ at the expected mass scales \textit{a posteriori}, although the posterior distribution for \gammahightwo~is bimodal and also supports a small over-density at higher masses~\citep[i.e., it snaps to the component masses of GW190521;][]{GW190521}. 

\MultiPDB~(bottom row of Fig.~\ref{fig:results}) builds upon \PDB~by considering a model that explicitly adds two additional Gaussian peaks: (\peakone, \sigone) and (\peaktwo, \sigtwo) in Fig.~\ref{fig:params}.
\MultiPDB~consistently finds over-densities in $p_{1D}$ at both $\sim$\peakatten and $\sim$\peakatthirty.


\section{Joint posteriors conditioned on O3}
\label{sec:H0 constraints}

For each mass model, we sample from the joint posterior distribution for both the parameters describing the mass distribution and the Hubble relation. 
Fig.~\ref{fig:results} shows the median and 90\% symmetric credible regions for $p_{1D}$ as a function of mass along with the marginal posterior distributions for $H_0$.

Generally, we find posteriors with the expected behaviour that are consistent with previous results \citep{lvkcosmoO3}.
All models prefer a dip between \lowerMGlow--\lowerMGhigh.
All models also find a local maximum in $p_{1D}$ around $\sim$ \peakatten, although this manifests as a more pronounced ``peak'' in models which support multiple peaks (\DoubleDip~and \MultiPDB).
All models also consistently find that the power-law steepens at higher masses through a roll-off above \shelf. 
Finally, all models that can support a local maximum near $\sim$ \peakatthirty~find one \textit{a posteriori}.
Models with Gaussian peaks call this feature \peaktwo, and it corresponds to \gammahightwo~in \DoubleDip~(the upper edge of the second notch).

Symmetric posterior credible regions for all other values included in our fits can be found in Table~\ref{tab:posteriors}.
We report their median values along with their 90\% symmetric credible region.

Additionally, there are qualitative similarities in the marginal posteriors for $H_0$.
All models primarily disfavour large values of $H_0$ \textit{a posteriori}, which would correspond to large detector-frame masses given a fixed source-frame mass distribution.
The $H_0$ posteriors obtained with \PDB, \PDBP, and \MultiPDB~closely resemble the results from \cite{lvkcosmoO3}, which assumed a source-frame primary-mass distribution consisting of a mixture of a single (unbroken) power-law and Gaussian peak.
Interestingly, the posterior obtained with \DoubleDip~has a local maximum near the values reported by the Planck and SH0ES collaborations \citep{planck2018, Riess_2021}. However, the posterior is very wide and remains consistent with the other mass models.

It is clear, then, that current catalogues of CBCs are not competitive with other estimates of $H_0$.
However, it is also clear that some cosmological information is encoded within the inference.
We now examine exactly how information about $H_0$ manifests through correlations with different features in the source-frame mass distribution.


\section{Which features carry cosmological information?}
\label{sec:best_feature}

Our mass models support a wide range of features, including peaks, notches, and roll-offs.
Upon first inspection, it may not be immediately clear which of these features would be most useful when attempting to infer the Hubble relation.
We attempt to identify those features by examining the Pearson correlation coefficients ($r$) between $H_0$ and features in $p_{1D}$ \textit{a posteriori} (i.e., correlations induced by conditioning on the observed data). 
Table~\ref{tab:features} reports the correlation coefficients between $H_0$ and several statistics representing these features.

\begin{table*}
\centering
    \caption{
        Pearson correlation coefficients ($r$) between features in the mass distribution and $H_0$ from joint posterior distributions conditioned on CBCs from O3. 
        Individual model parameters are described in Sec.~\ref{sec:mass distributions} and Fig.~\ref{fig:params}.
        Model-independent summary statistics are described in Sec.~\ref{sec:nonparametric features}.
    }
    \label{tab:features}
    \begin{tabular}{l |c| ccc | cccc | cc}
        \toprule
            & \gammahighone & \peakone & \peakhatone & \sighatone & \peaktwo & $(\peaktwo | \sigtwo < $\cutpeak$\,M_{\odot})$ & \peakhattwo & \sighattwo & \mmax & \mninenine \\
        \midrule
            \PDB
                & \rPDBgammahighone & \rPDBmupeakone & \rPDBmuhatone & +\rPDBsigmahatone & \rPDBmupeaktwo & \rPDBmupeaktwoconditioned & \rPDBmuhattwo & \rPDBsigmahattwo & \rPDBmmax & \rPDBmninenine \\
            \PDBP
                & \rPDBPgammahighone & \rPDBPmupeakone & \rPDBPmuhatone & +\rPDBPsigmahatone & \rPDBPmupeaktwo & \rPDBPmupeaktwoconditioned & \rPDBPmuhattwo & \rPDBPsigmahattwo & \rPDBPmmax & \rPDBPmninenine \\
            \DoubleDip
                & \rDoubleDipgammahighone & \rDoubleDipmupeakone & \rDoubleDipmuhatone & \rDoubleDipsigmahatone & \rDoubleDipmupeaktwo & \rDoubleDipmupeaktwoconditioned & \rDoubleDipmuhattwo & \rDoubleDipsigmahattwo & \rDoubleDipmmax & \rDoubleDipmninenine \\
            \MultiPDB
                & +\rMultiPDBgammahighone & \rMultiPDBmupeakone & \rMultiPDBmuhatone & \rMultiPDBsigmahatone & \rMultiPDBmupeaktwo & \rMultiPDBmupeaktwoconditioned & \rMultiPDBmuhattwo & \rMultiPDBsigmahattwo & \rMultiPDBmmax & \rMultiPDBmninenine \\
        \bottomrule
    \end{tabular}
\end{table*}

We currently expect high-mass (BBH) features to carry more cosmological information than BNS masses.
This is because BBH mergers are detectable to much larger redshifts that BNS.\footnote{\cite{Ezquiaga_2022} claim that the lower mass gap will eventually dominate the constraint with next-generation GW detectors given the expected increase in the BNS detection rate.}
As such, more BBHs are detected within current catalogues, and cosmological effects are more apparent within BBHs.
We begin by examining the correlations between $H_0$ and a few parameters of each mass distribution in Sec.~\ref{sec:parametric features}

As we will see, it quickly becomes apparent that relying on specific parameters of individual models may be difficult to scale, as models with different parametrizations can nevertheless capture the same behaviour.
As such, we additionally consider several model-independent summary statistics derived from $p_{1D}$ and study how they correlate with $H_0$ in Sec.~\ref{sec:nonparametric features}.
These statistics can be extended to any mass model, even those without concise functional forms \citep{Callister_2024, Edelman_2023, ray2024searchingbinaryblackhole, farah2024needknowastrophysicsfreegravitationalwave}.


\subsection{Parametric descriptions of the mass distribution}
\label{sec:parametric features}

Examining the behaviour of the high-mass end of the distributions in Fig.~\ref{fig:results}, we might expect either the shelf created by the roll-off around \shelf~or local maximum at $\sim$ \peakatthirty~to carry cosmological information.
In general, $H_0$ may correlate with many parameters, such as \mmax~(roll-off at high masses), \peakone~(overdensity near \peakatten), \peaktwo (overdensity near \peakatthirty).
Some of these are listed in Table~\ref{tab:features}. 

To begin, we examine the roll-off at high masses, which was previously identified as a useful feature in~\citet{Farr_2019}.
Even if the true mass distribution does not have any peaks, like \PDB, the roll-off at high masses still carries cosmological information.
That is, we do not need a peak to constrain the Hubble relation.

\begin{figure*}
    \includegraphics[width=0.5\textwidth] {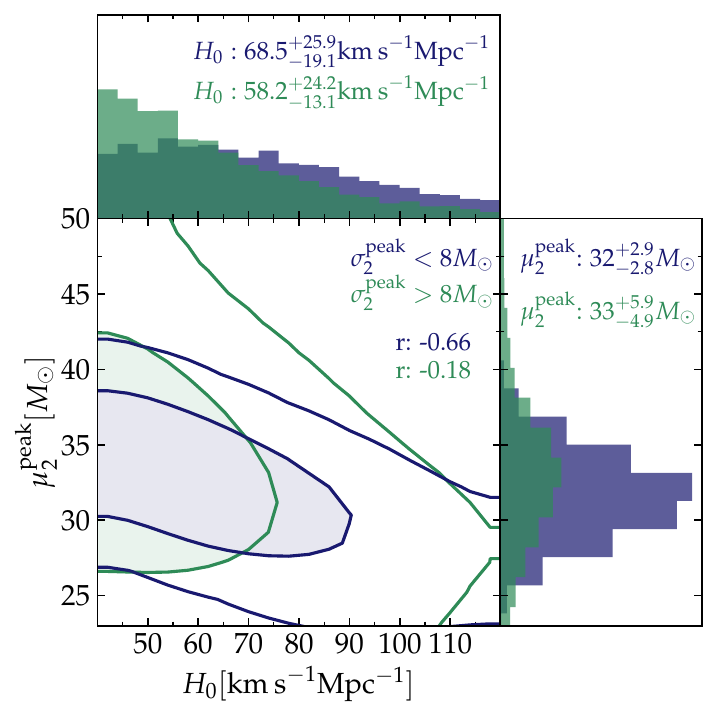}
    \includegraphics[width=0.5\textwidth] {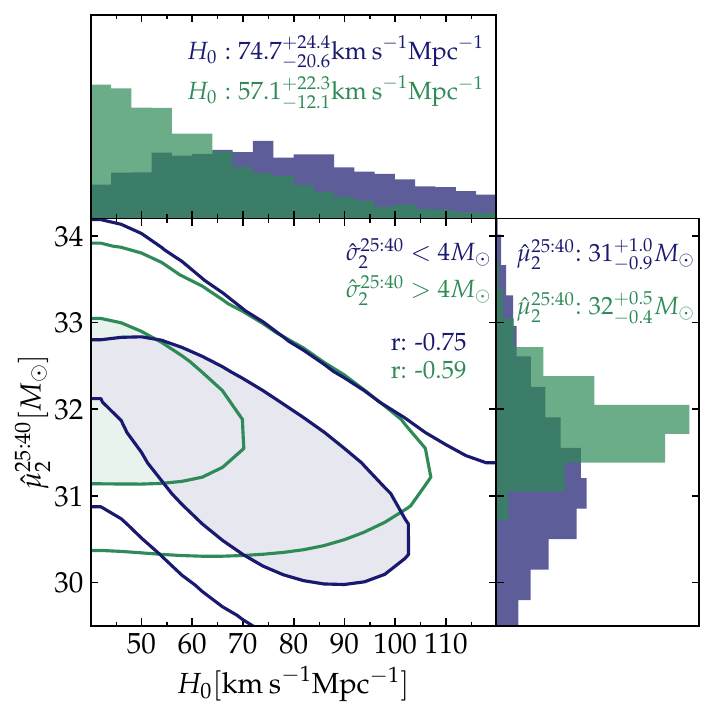}
    \caption{
        (\emph{left}) Joint and marginal posterior distributions for \peaktwo~and $H_0$ conditioned on the presence of wide (\emph{green}, $\sigtwo \geq $\cutpeak$\,\rm M_\odot$) and narrow (\emph{blue}, $\sigtwo \leq$\cutpeak$\,\rm M_\odot$) peaks with \MultiPDB.
        Contours in the joint distributions show the 50\% and 90\% highest-probability-density credible regions.
        Median and 90\% symmetric credible regions are shown in the marginal distributions, and the Pearson correlation coefficients ($r$) are shown in the joint distribution.
        (\emph{right}) Analogous model-independent summary statistics derived from $p_{1D}$ (\peakhattwo~and \sighattwo) for \MultiPDB. 
        We apply different thresholds on \sigtwo~and \sighattwo~based on their one-dimensional posterior medians.
    }
    \label{fig:sharp_peaks}
\end{figure*}

However, Table~\ref{tab:features} also shows that when a model supports a peak near \peakatthirty~, the location of that peak always correlates more strongly with $H_0$ than \mmax.
This is also apparent in Figs. 5 and 13 of~\citet{lvkcosmoO3}.
As such, while the presence of a peak is not necessary for spectral siren cosmology, it is helpful.
Additionally, we consider which aspects of the peak help it carry cosmological information in Fig.~\ref{fig:sharp_peaks} using \MultiPDB.
In line with predictions from~\citet{taylor2012cosmology}, we find that wide peaks (\sigtwo$ \geq$ \cutpeak $\,\rm M_\odot$) yield small correlation coefficients \textit{a posteriori}, even smaller than what is observed between $H_0$ and \mmax.
However, relatively sharp peaks ($\sigtwo \leq$\cutpeak$\,\rm M_\odot$) yield strong correlations between $H_0$ and \peaktwo.
Interestingly, conditioning on a sharp peak significantly shifts the maximum of the marginal posterior for $H_0$ much closer to $70 \mathrm{km\:s}^{-1}\mathrm{Mpc}^{-1}$.

Finally, Table~\ref{tab:features} does not report a correlation between \peaktwo~and $H_0$ for \DoubleDip~because that model does not contain that parameter.
Interestingly, \DoubleDip~does have a few parameters that moderately correlate with $H_0$.
The lower and upper edges of the high-mass dip (\gammalowtwo~and \gammahightwo) both correlate with $H_0$ with $r=\rDoubleDipgammalowtwo$ and \rDoubleDipgammahightwo, respectively~\footnote{Simpsons paradox~\citep{Simpson:1951} indicates that correlations between variates observed in the whole population can change when subsets of samples are examined separately. The \gammahightwo correlations with $H_0$ demonstrate this behaviour (since \gammahightwo is bimodal). Both sub-populations independently anti-correlate with $H_0$ (we have confirmed this). However, when combined, their correlations appear positively correlated.}.
Additionally, \mmax~and \etamax, which trace the location and steepness of the fall-off at masses above $\sim40 M_{\odot}$, also correlate with $H_0$ ($r =  \rDoubleDipmmax$ and \rDoubleDipetamax, respectively).
However, none of these parameters correlate as strongly with $H_0$ as the parameters describing peaks in other models (e.g., \peaktwo~in \MultiPDB).
That is, even though \DoubleDip~can reproduce qualitatively similar features as the other models in Fig.~\ref{fig:results}, it lacks a single parameter that concisely captures the overall behaviour of $p_{1D}$ near $\sim 30\,\rm M_\odot$.
This makes it clear that relying on correlations between individual parameters and $H_0$ may not be a robust way to describe the information contained within $p_{1D}$ and suggests the need for model-independent summary statistics, which we explore in Sec.~\ref{sec:nonparametric features}.

\vspace{0.1in}
\subsection{Model-independent summary statistics derived from the mass distribution}
\label{sec:nonparametric features}

So far, our analysis has focused on specific parameters in individual models.
However, this may not generalize well, particularly for models that do not have concise parametric representations~\citep{farah2024needknowastrophysicsfreegravitationalwave, Callister_2024, ray2024searchingbinaryblackhole}.
Instead, we now characterize general features in the mass distribution with model-independent summary statistics and then consider their correlations with $H_0$.

We introduce several such statistics.
Each is derived from $p_{1D}$, and therefore not directly applicable to the joint distribution $p(m_{1s}, m_{2s}|\Lambda)$ that may be constrained by other analyses.
However, we believe they nevertheless capture important behaviour and can be generalized to the joint distribution straightforwardly, particularly as GW catalogues seem to suggest a relatively strong preference for equal-mass binaries within the astrophysical distribution \citep{Fishbach_2020, theligoscientificcollaboration2022populationmergingcompactbinaries}.

First, we compute moments concerning $p_{1D}$ over restricted ranges of mass.
Specifically, we consider the mean and variance within a range of masses:
\begin{align}
    \hat{\mu}(X, Y) & = \int_{X}^{Y} m q(m; X,Y) dm \\
    \hat{\sigma}^2(X:Y) & = \int_{X}^{Y} \left[m - \hat{\mu}(X,Y)\right]^2 q(m; X,Y) dm
\end{align}
where
\begin{equation}\label{eq:nonparam_estimators}
    q(m; X,Y) \equiv \frac{p_{1D}(m|\Lambda)}{\int_{X}^{Y} p_{1D}(m|\Lambda)dm} \, \Theta(X \leq m \leq Y)
\end{equation}
Table~\ref{tab:features} reports the correlations between $H_0$ and
\begin{align}
    \peakhatone  & \equiv \hat{\mu}(7\,\rm M_\odot, 11\,\rm M_\odot) \\
    \sighatone & \equiv \hat{\sigma}(7\,\rm M_\odot, 11\,\rm M_\odot)
\end{align}
as well as
\begin{align}
    \peakhattwo    & \equiv \hat{\mu}(25\,\rm M_\odot, 40\,\rm M_\odot) \\
    \sighattwo & \equiv \hat{\sigma}(25\,\rm M_\odot, 40\,\rm M_\odot)
\end{align}
which are intended to model the peaks observed with \DoubleDip~and \MultiPDB~near $\sim$ \peakatten~and $\sim$ \peakatthirty, respectively.

The precise behaviour of these statistics depends somewhat on the integration bounds, which can complicate their interpretation.
Put simply, \peakhattwo~does not trace \peaktwo~exactly.
Instead, it acts as an accumulation of information within the range that \peaktwo~operates.
Similarly, \sighattwo~will not perfectly trace \sigtwo.
We do not expect our correlations between these statistics and $H_0$ to greatly change with different integral bounds and have confirmed that changing them does not alter our conclusions.

We also consider the $99^\mathrm{th}$ percentile (\mninenine) of $p_{1D}$ as a model-independent proxy for the location of the roll-off at high masses, analogous to \mmax~in Sec.~\ref{sec:parametric features}.
Again, we find the same general trends for different percentiles above $\sim 95\%$.

Interestingly, Table~\ref{tab:features} shows that the model-independent statistics (almost) always correlate with $H_0$ more strongly than their parametric analogs.
At first glance, this may be surprising, as one might expect individual model parameters to trace features in the mass distribution better than \textit{ad hoc} summary statistics.
However, as we see in Fig~\ref{fig:sharp_peaks}, the correlation of some parameters with $H_0$ can depend on the values taken by other parameters.
When \sigtwo~is large, \peaktwo correlates poorly with $H_0$ (r = 0.18). 
In this scenario, Table~\ref{tab:features} shows that \mmax~is a better correlator.
Conversely, the summary statistics can capture the relevant behaviour within the mass distribution regardless of the behaviour of individual parameters (i.e., when \sigtwo~is large and \peaktwo~no longer correlates strongly with $H_0$, \peakhattwo~can still pick up on the roll-off associated with \mmax).

Fundamentally, then, it is likely that our summary statistics correlate more strongly with $H_0$ because they are sensitive to the overall shape of $p_{1D}$.
Individual model parameters may be degenerate, meaning that the same approximate shape of $p_{1D}$ may be obtained with several different parameter combinations.
A degeneracy between parameter A and parameter B may weaken their individual correlations with $H_0$. 
Out summary statistics, which may not be closely tied to an individual parameter, allow us to study the relationship between $H_0$ and the overall shape of $P_{1D}$.

Fig.~\ref{fig:sharp_peaks} further demonstrates that our model-independent statistics capture the expected behaviour.
The correlation between \peakhattwo~and $H_0$ also improves when we condition on small \sighattwo~(i.e., a narrow feature).
This observation, combined with the fact that \peakhattwo~consistently has one of the largest correlation coefficients in Table~\ref{tab:features} across all models, strongly suggests that the bulk of the cosmological information in current GW catalogues is carried by binaries with source-frame masses covered by \peakhattwo~(i.e., 25-40$\,\rm M_\odot$).

This makes sense.
High-mass binaries are detected at larger distances (and redshifts), meaning that they can have a larger cosmological imprint on their detector-frame masses.
We also tend to detect more of them.
However, the properties of the features themselves also play a role.
Fig~\ref{fig:sharp_peaks} shows that sharper peaks correlate with $H_0$ more strongly than broad peaks.
Furthermore, even though systems from the low-mass feature have a smaller detection horizon, the peak near $\sim$ \peakatten~may correlate with $H_0$ nearly as well as the peak near $\sim$ \peakatthirty~because it is sharper.

We can also understand the signs of the correlations between the summary statistics in Table~\ref{tab:features} and $H_0$ as follows.
The correlations observed \textit{a posteriori} are driven by the different combinations of source-frame mass and $H_0$ that can predict the same detector-frame mass, which for low $z$ is approximately
\begin{equation}\label{eq:dimensional analysis}
    m_d \approx \left(1 + \frac{H_0 D_L}{c}\right) m_s
\end{equation}
The location of peaks (e.g., \peakhatone~and \peakhattwo) are anti-correlated with $H_0$ because increasing either peak will tend to increase $m_s$, which in turn must be compensated by a decrease in $H_0$.
Table~\ref{tab:features} shows that \gammahighone~(upper edge of the lower notch) is also usually anti-correlated with $H_0$, for the same reason.

The behaviour of the summary statistics for the width of the peaks (\sighatone~and \sighattwo) can be more complicated, though.
In the presence of a peak, we observe that $\hat{\sigma}$ is anti-correlated with $H_0$.
This is because larger $\hat{\sigma}$ produce a wider astrophysical source-frame mass distribution and, because more massive binaries are easier to detect than less massive binaries, this shifts the mean source-frame mass in the detected distribution to larger values.
As such, increasing $\hat\sigma$ produces a similar effect to increasing $\hat\mu$, which requires $H_0$ to decrease to maintain the same $m_d$ in Eq.~\ref{eq:dimensional analysis}.

Interestingly though, in the absence of a peak, \sighatone~is positively correlated with $H_0$.  
This is the opposite of the behaviour \sighattwo demonstrates.
We understand this as the effect of a strong anti-correlation between \gammahighone~and \sighatone~in the absence of a peak near $\sim 9\,\rm M_\odot$.
That is, when \gammahighone~increases, it tends to cut out part of the mass distribution within the range spanned by \sighatone, which produces a narrower distribution and a smaller value of $\sighatone$ (see posterior credible regions for \gammahighone~in Table~\ref{tab:posteriors}).
Therefore, we observe a positive correlation between \sighatone~and $H_0$ because \sighatone~is anticorrelated with \gammahighone, which in turn is anticorrelated with $H_0$.

\begin{figure}
    \begin{center}
    \includegraphics[width=0.90\columnwidth, clip=True, trim=0.0cm 0.5cm 0.0cm 0.0cm]{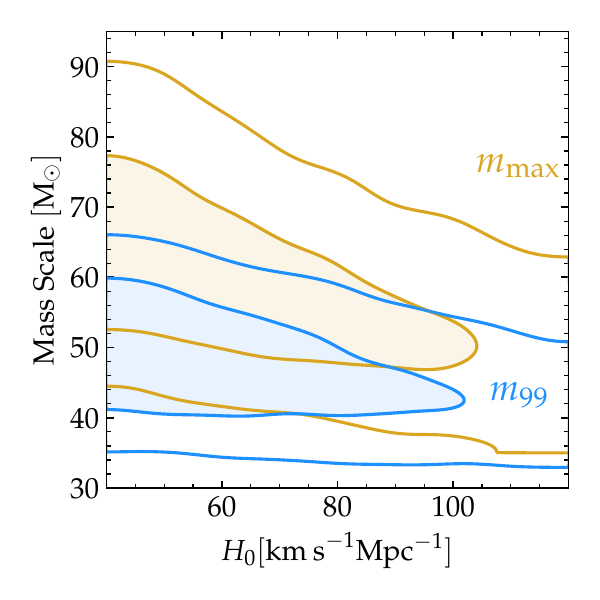}
    \end{center}
    \caption{
        Joint posterior distributions for $H_0$ and \mmax, \mninenine~obtained with \PDB.
        Contours denote the 50\% and 90\% highest-probability-density credible regions.
        Both mass scales trace the location of the roll-off at high masses. Even for models without prominent peaks, cosmological information is still encoded in the mass distribution. 
    }
    \label{fig:not_peak}
\end{figure}

Finally, we again note that, while peaks are helpful for cosmological constraints, they are not necessary.
Fig.~\ref{fig:not_peak} compares the joint posteriors between $H_0$ and \mmax, \mninenine~for \PDB, which, unlike \PDBP, \DoubleDip, and \MultiPDB, does not support a peak near $\sim$ \peakatthirty.
Both $m_\mathrm{max}$ and $m_{99}$ correlate with $H_0$ comparably.
Broadly similar behaviour is seen across all models, and similar behaviour manifests in \peakhatone~and \sighatone in \PDB~and \PDBP.
Again, this is likely because \mninenine~is sensitive to the overall shape of $p_{1D}$, whereas \mmax~may not be important if there is a large peak in the same mass range.
This often seems to be the case in Fig.~\ref{fig:results}.

\section{Independent correlations between multiple features and $H_0$}
\label{sec:other_features}

\begin{figure}
    \begin{center}
    \includegraphics[width=0.90\columnwidth, clip=True, trim=0.0cm 0.1cm 0.0cm 0.0cm]{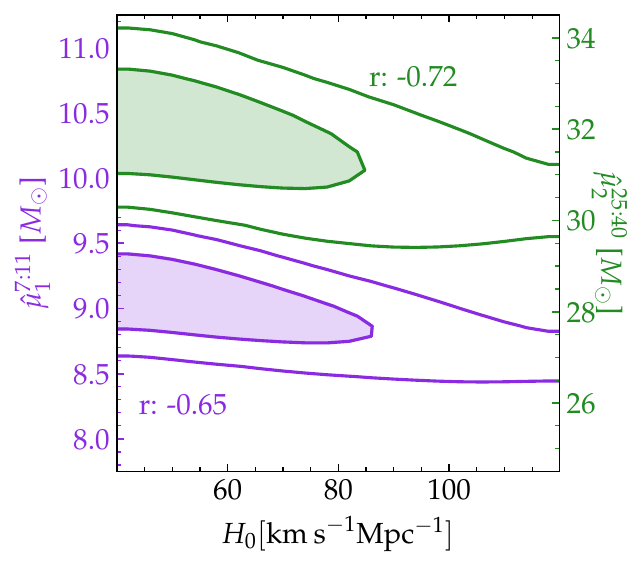}
    \end{center}
    \caption{
        Joint posteriors between $H_0$ and \peakhatone~(\emph{purple}), \peakhattwo~(\emph{green}).
        Contours denote the 50\% and 90\% highest-probability density credible regions. The estimator functions defined in Eq.~\ref{eq:nonparam_estimators} are applied to \MultiPDB. They correlate similarly for both local overdense regions. 
    }
    \label{fig:mu_hat_corr}
\end{figure}

As seen in Table \ref{tab:features}, \peakhatone~can correlate with $H_0$ nearly as strongly as \peakhattwo, particularly for models that support multiple peaks (\DoubleDip~and \MultiPDB).
Fig.~\ref{fig:mu_hat_corr} shows the joint posteriors for \MultiPDB.
This would appear to be great news for spectral siren cosmology, as one of the main advantages of the method is that multiple features in the mass distribution can be used to constrain the Hubble relation at the same time.

However, it is not immediately clear whether these features independently correlate with $H_0$.
That is, it could be the case that \peakhatone~appears to correlate with $H_0$ only because it correlates with \peakhattwo~as they are both related to the shape of the overall mass distribution.
Conditioning the posterior on the observed catalogue can easily introduce correlations like this.
Put another way, \peakhatone~could correlate with \peakhattwo~even if we fixed $H_0$.
In this way, information about $H_0$ could pass from one feature to another.
Therefore, it could be the case that the correlation between \peakhatone~and $H_0$ observed in Table~\ref{tab:features} and Fig.~\ref{fig:mu_hat_corr} are only present because \peakhatone~separately correlates with \peakhattwo~which in turn correlates with $H_0$. We will disprove this by demonstrating that for \peakhatone~to be useful in spectral siren measurements, it must directly correlate with $H_0$ independently of \peakhattwo.

\begin{figure*}
    \begin{center}
    \includegraphics[width=0.80\textwidth, clip=True, trim=0.0cm 0.5cm 0.0cm 0.0cm]{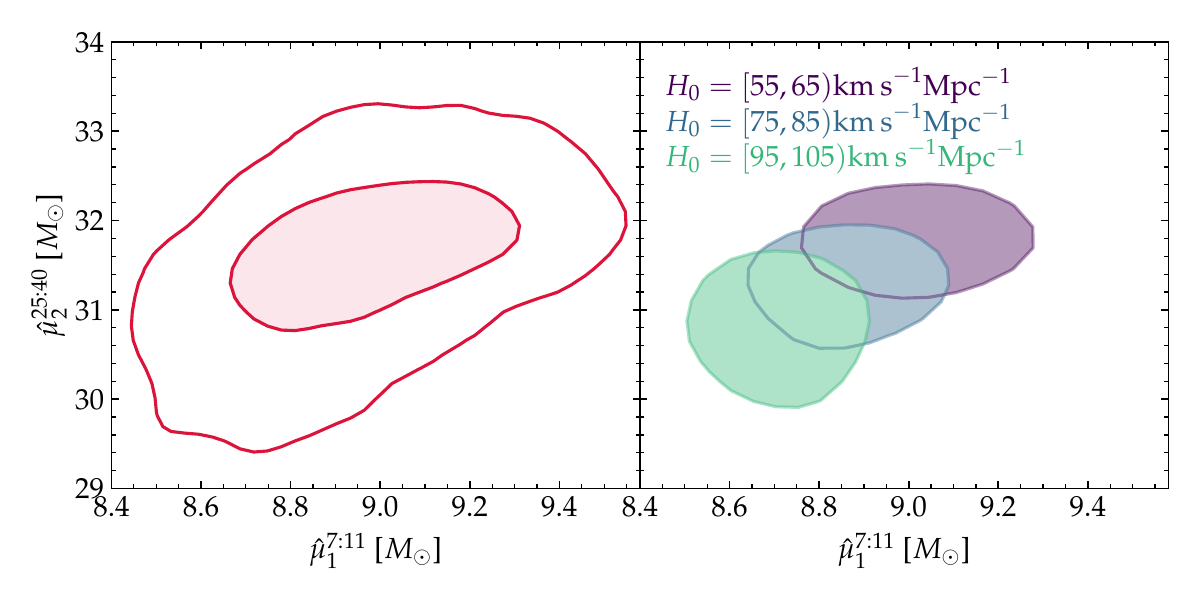}
    \end{center}
    \caption{
        (\emph{left}) Joint posterior distribution over \peakhatone~and \peakhattwo~in \MultiPDB. 
        Contours denote the 50\% and 90\% highest-probability-density credible regions.
        (\emph{right}) 50\% posterior credible regions from the joint posterior conditioned on different ranges of $H_0$: 55-65 (\emph{purple}), 75-85 (\emph{blue}), and 95-105 (\emph{teal}) $\mathrm{km}/\mathrm{s}/\mathrm{Mpc}$.
        We omit the 90\% credible regions for clarity, as they significantly overlap.
    }
    \label{fig:mu_hat_conditioned}
\end{figure*}

Another way to phrase this is whether \peakhatone~and \peakhattwo~are conditionally independent given $H_0$ ($\peakhatone \perp \peakhattwo \, | \, H_0$).
Fig.~\ref{fig:mu_hat_conditioned} directly addresses this by plotting the joint posterior distributions for \peakhatone~and \peakhattwo~conditioned on several (small) ranges of $H_0$.
We see that \peakhatone~and \peakhattwo~are correlated in their joint posterior, but essentially all of the correlation is due to their separate correlations with $H_0$.
In fact, \peakhatone~and \peakhattwo~are almost completely uncorrelated when we condition on $H_0$, but their joint posterior shifts to larger values as we increase $H_0$.\footnote{This is another manifestation of Simpson's paradox}

To quantify how much of the correlation between \peakhatone~and \peakhattwo~is due to their separate correlations with $H_0$, we consider a simple model in which \peakhatone~and \peakhattwo~separately depend linearly on $H_0$.\footnote{There is no compelling reason to believe the dependence is this simple; we only use this model to motivate more general expressions.}
\begin{align}
    \peakhatone & = (a H_0 + b) + \delta \\
    \peakhattwo & = (\alpha H_0 + \beta) + \epsilon
\end{align}
where $\delta$ and $\epsilon$ are drawn from a joint distribution that does not depend on $H_0$.
If we additionally assume $H_0 \sim p(H_0)$, it is straightforward to show that
\begin{equation}
    \mathrm{C}[\peakhatone, \peakhattwo] = \frac{\mathrm{C}[\peakhatone, H_0]\mathrm{C}[\peakhattwo, H_0]}{\mathrm{V}[H_0]} + \mathrm{C}[\delta, \epsilon]
\end{equation}
where $\mathrm{C}[x,y]$ denotes the covariance between $x$ and $y$ and $\mathrm{V}[x] = \mathrm{C}[x,x]$ is the variance of $x$.
The first term represents the covariance induced by the fact that both \peakhatone and \peakhattwo depend on the same variable ($H_0$), and the second term corresponds to the ``extra'' covariance between the two that would exist even at fixed $H_0$: $\mathrm{C}[\peakhatone, \peakhattwo | H_0]$.
It is natural to compare the size of these terms, and we define the ratio
\begin{equation} \label{eq:relativecosmo}
    \mathcal{R} = \left| \frac{\mathrm{C}[\peakhatone, H_0] \mathrm{C}[\peakhattwo, H_0]}{\mathrm{C}[\peakhatone, \peakhattwo] \mathrm{V}[H_0] - \mathrm{C}[\peakhatone, H_0]\mathrm{C}[\peakhattwo, H_0]} \right|
\end{equation}
If $\mathcal{R} \gg 1$, then the vast majority of the correlation between \peakhatone~and \peakhattwo~is due to their separate dependence on $H_0$, which is the ideal situation for a spectral sirens constraint.
Table~\ref{tab:reeds_corrs} summarizes these results, and we find that the model-independent estimates' covariance is indeed almost entirely driven by their separate correlations with $H_0$.
The lone exception to this is \PDB, which can be explained by the poor correlation between \peakhatone~and \peakhattwo.
Even then, the correlation from the separate dependence on $H_0$ is almost as large as all other factors combined.

Although we restrict our analysis to \peakhatone and \peakhattwo, a similar analysis can be applied to other estimators. We leave such studies for future work.
\begin{table}
    \centering
    \caption{
        Pearson correlation coefficients ($r$) for \peakhatone~and \peakhattwo~along with the ratio of covariances from their joint dependence on $H_0$ and all other factors ($\mathcal{R}$, Eq.~\ref{eq:relativecosmo}), ideally $R\gg1$.
        All models with significant correlations between \peakhatone~and \peakhattwo~also have large $\mathcal{R}$.
    }
    \label{tab:reeds_corrs}
    \begin{tabular}{l | ccc}
        \toprule
             & $r$ & $\mathcal{R}$ \\
        \midrule
            \PDB
                & \statPDBr & \statPDBmathcalR \\
            \PDBP
                & \statPDBPr & \statPDBPmathcalR \\
            \DoubleDip
                & \statDoubleDipr & \statDoubleDipmathcalR \\
            \MultiPDB
                & \statMultiPDBr & \statMultiPDBmathcalR \\
        \bottomrule
    \end{tabular}
\end{table}
 

\section{Discussion}
\label{sec:discussion}

We have shown that the observed mass distribution, as inferred with CBCs at all mass scales from O3, appears to contain multiple features that separately correlate strongly with $H_0$. 
Future analysis can use this work to further improve spectral siren constraints. 
However, we make several simplifying assumptions in our analysis that may require further study. 

First, although we considered several variations of the one-dimensional latent mass distribution $p_{1D}$, we only considered a single functional form for a mass-dependent pairing function in Eq.~\ref{eq:joint mass}.
We also assume that $m_{1s}$ and $m_{2s}$ are drawn from the same $p_{1D}$.
Alternatives have been proposed in the literature.
\cite{farah2023kindcomparingbigsmall} looked at whether there is evidence that $m_{1s}$ and $m_{2s}$ are drawn from the same underlying 1D mass distributions. They found that current data slightly favours our model choice.
\cite{Fishbach_2020} and \cite{farah2022bridging} examined different pairing functions with earlier GW catalogues; both found that our pairing function is a reasonable description of the data.
It is also similar to the common model assumption that $p(m_{2s}|m_{1s}) \propto m_{2s}^{\beta}$ \citep{farah2023kindcomparingbigsmall}.

In general, model misspecification of this kind is a persistent concern in any inference that assumes a specific functional form for the mass distribution. \cite{pierra2024study} shows that mass model choices may bias $H_0$ measurements.
We believe our parametric models provide a reasonable fit to the current GW catalogue, particularly for BBH masses where a strong preference for equal-mass binaries has been observed \citep{Fishbach_2020}.
Recent non-parametric mass models also show that the source-frame mass distribution we use is realistic and matches the observed data \citep{Rinaldi_2021, Tiwari_2021, farah2024needknowastrophysicsfreegravitationalwave, ray2024searchingbinaryblackhole, Callister_2024}.
What's more, just as models with peaks may be better able to constrain $H_0$ than models without peaks, models with more complicated pairing functions (that depend on scales in the source frame) should only improve the ability of spectral sirens to constrain $H_0$.
In this respect, our assumption may in fact be conservative.

In addition to our mass distribution, our analysis makes several assumptions about the behaviour of the distribution of merging binaries with redshift.
In particular, one may be concerned that we do not know the true redshift distribution $p(z|\Lambda)$ and that our assumption of a fixed redshift distribution (which tracks the star formation rate) is overly optimistic.
While additional uncertainty in $p(z|\Lambda)$ is likely to weaken posterior constraints on $H_0$, we do not expect it to completely spoil our ability to constrain $H_0$.
This may change the typical redshift at which we observe events, but it is unlikely to change the relationship between $m_d$ and $z$, which is what connects back to specific features in $p(m_{1s}, m_{2s}|\Lambda)$ upon which we base our measurement.
However, additional work is warranted to show that this is indeed the case.
\footnote{Single-event measurement uncertainties on $D_L$ can be broad, and different population priors may be able to significantly alter our posterior beliefs about individual events.}

The other significant assumption about the redshift dependence within our analysis is that the source-frame mass distribution $p(m_{1s}, m_{2s}|\Lambda)$ is independent of $z$.
In reality, this almost certainly is not the case.\footnote{Current observations do not rule out redshift evolution of the source-frame mass distribution, but it is not required to explain the data either \citep{Fishbach_2021}.}
It is known that metallicity correlates with redshift and, therefore, stars that form earlier in the universe form in more metal-poor environments \citep{Madau_2014}.
We expect the metallicity of the formation environment to impact the masses of the stellar remnants left behind \citep{Madau_2017}.
Additionally, multiple separate formation channels may have been active during different epochs of cosmic history.
If these formation channels preferentially produce different types of binaries, then, again, the source-frame mass distribution may depend on redshift~\citep{Ng:2020qpk, vanSon:2021zpk, Ye:2024ypm}.
Given that the location and shapes of features in the source-frame mass distribution may shift with redshift, some authors have raised the reasonable concern that this astrophysical evolution could be confused with cosmological redshift. Both could manifest as a dependence of the detector-frame mass distribution on the luminosity distance. 

While our assumption that the source-frame mass distribution is independent of redshift does not allow us to directly address this concern, we do provide evidence that it may not be a showstopper.
Specifically, while we may expect the source-frame mass distribution to evolve, we do not expect it to evolve in the same way at all mass scales.
However, a cosmological redshift will affect all mass scales in the same way. 

This is similar to how redshifts observed for both atomic and molecular transition lines are evidence for cosmological redshift rather than a conspiracy of changes in the underlying physics for each set of lines. Changes in the electron mass could affect atomic transitions in, e.g., Hydrogen, but molecular transitions in $H_2$ are controlled by the mass of the proton. It is more parsimonious to infer cosmological redshift than to contrive a model that changes the mass of both the electron and the proton simultaneously to mimic the observed behaviour.

In the same way, if multiple features in the source-frame mass distribution correlate separately with $H_0$ (as we have shown is the case in Sec.~\ref{sec:other_features}), we expect to be able to break the degeneracy between astrophysical and cosmological effects.
See \cite{farah2024needknowastrophysicsfreegravitationalwave} for an explicit demonstration of how this could work with simulated data.

So far, the features we identified are within the BBH portion of the mass distribution.
However, it is expected that the larger detection rate of BNS with next-generation GW detectors will eventually drive spectral siren constraints~\citep{Ezquiaga_2022}.
BNS and NSBH will also provide other tracers of the Hubble relation, as tidal effects in the GW waveform will not redshift and can be used as an alternate measurement of the source-frame masses.
In general, any binary property that does not redshift but is correlated with the source-frame masses, like NS maximum mass or component spins, could be used in this way~\citep{Chatterjee_2021, Chen_2024, Messenger_2012, Mukherjee_2022, Ghosh_2022, ghosh2024jointinferencepopulationcosmology}.
However, it is difficult to measure tides and spins in the current catalogue of CBCs, so it is not clear how much of a near-term improvement these additional features could provide.
While we leave further investigations to future work, we note that additional information of this type should only improve our ability to measure $H_0$ from GW catalogues without electromagnetic counterparts: the more independent features, the easier it is to disentangle astrophysical evolution from cosmology. 

Our results show that it is, in fact, quite difficult to completely remove cosmological information from the observed distribution of CBC properties.
When we condition our hierarchical model on observed data, any features in the source-frame mass distribution, such as dips, gaps, peaks, and roll-offs, correlate with $H_0$ (and the full Hubble relation).
We identify several robust, model-independent features and show that they correlate strongly with $H_0$ across a range of models.
We also show that these features correlate with $H_0$ independently, which is the best-case scenario for spectral siren measurements.
Looking ahead, even in the presence of redshift evolution in the source-frame mass distribution, current data suggests we live in a universe in which spectral sirens can provide an accurate estimate of $H_0$.


\section*{Acknowledgements}
We sincerely thank Aditya Vijaykumar, José Ezquiaga, Maya Fishbach, Amanda Farah, and Phillipe Landry for the useful discussions throughout the project. 
U.M. and R.E. are supported by the Natural Sciences \& Engineering Research Council of Canada (NSERC) through a Discovery Grant (RGPIN-2023-03346).
This material is based upon work supported by NSF's LIGO Laboratory which is a major facility fully funded by the National Science Foundation. 


\software{
\texttt{numpy}~\citep{numpy}, \texttt{pandas}~\citep{reback2020pandas}, \texttt{numpyro}~\citep{phan2019composable, bingham2019pyro}, \texttt{jax}~\citep{jax2018github}, \texttt{matplotlib}~\citep{Hunter:2007}, \texttt{scipy}~\citep{2020SciPy-NMeth}.
}


\appendix


\section{adding vs. multiplying to create additional features}
\label{sec:pdb_p}

As mentioned in Sec.~\ref{sec:mass distributions}, there are several nearly equivalent methods for adding additional peaks to the mass distribution.
We focus on multiplicative filters that can either add a peak or remove a notch based on the sign of their amplitude parameter (see Appendix~\ref{sec:pop_models}).
However, it is also common to add a peak as an additional component within a mixture model.
We consider such a mixture for $p_{1D}$ by summing \PDB~and a Gaussian peak (\PDBpP; Eq.~\ref{eq:PDBpP}).

\begin{figure*}\label{fig:pdbp_model_compare}
    \centering
    \includegraphics[width=0.9\textwidth, clip=True, trim=0.0cm 0.25cm 0.0cm 0.75cm]{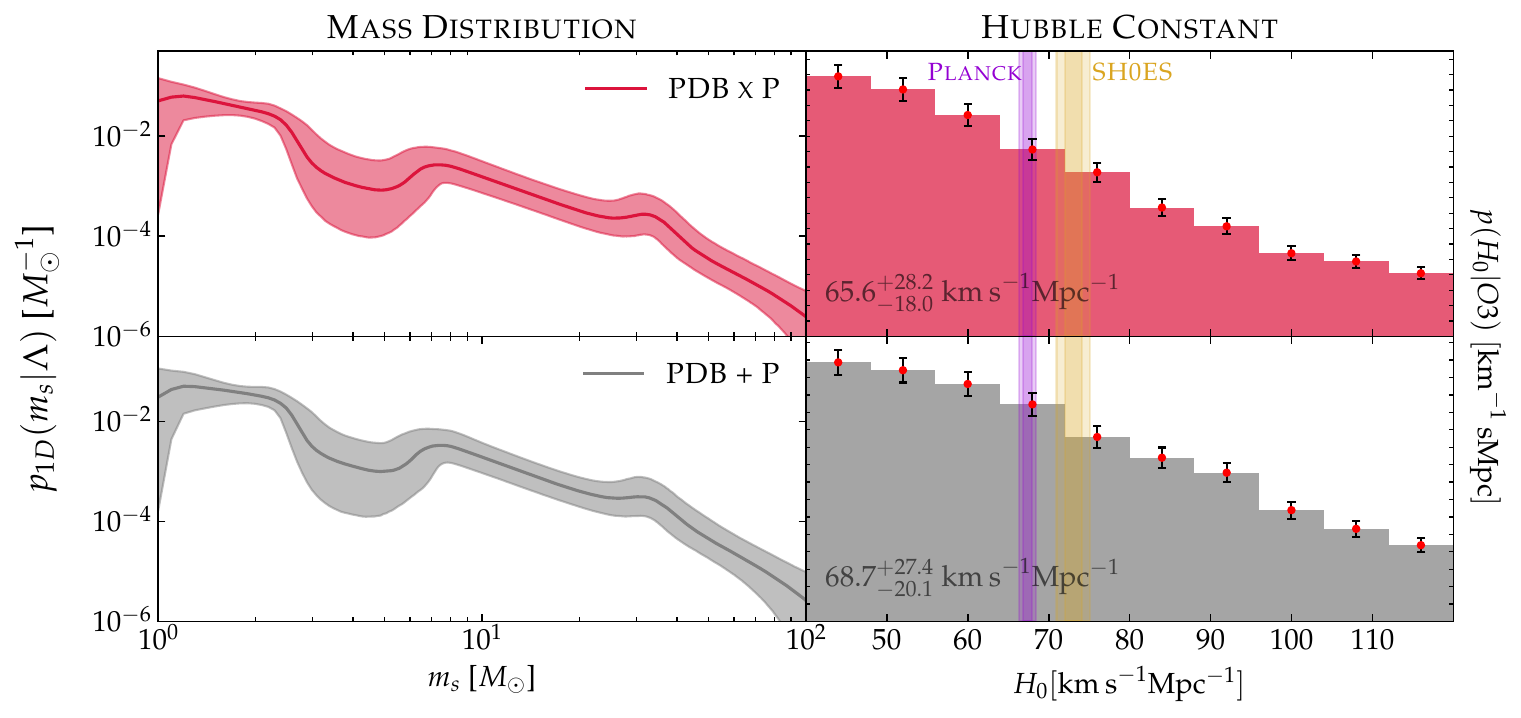}
    \caption{
        Similar to Fig.~\ref{fig:results}, we compare two different extensions of \PDB, each with a single peak: \PDBP~(Eq.~\ref{eq:PDBP}) and \PDBpP~(Eq.~\ref{eq:PDBpP}).
    }
    \label{fig:PDBP vs PDBpP}
\end{figure*}

We find quantitatively similar results with \PDBP~and \PDBpP~(Fig.~\ref{fig:PDBP vs PDBpP}).
Both marginal posteriors for $H_0$ yield similar constraints, the credible regions for $p_{1D}$ show similar features, and features in each mass model show similar correlations with $H_0$.

This suggests that, as expected, a model's behaviour does not strongly depend on the precise implementation of additional peaks, and analysts should choose whichever implementation is easiest to control.
For example, one does not need to normalize $p_{1D}$ within each likelihood call if additional features are included as multiplicative factors (the normalization cancels term-by-term between $\mathcal{Z}_i$ and $\mathcal{E}$ in Eq.~\ref{eq:likelihood}).
However, it may be necessary (and expensive!) to numerically normalize each mixture model component within each likelihood evaluation to obtain interpretable mixing fractions.  

\section{Details of the Population Model}
\label{sec:pop_models}

We define the population model as a product of distributions over the binary's source-frame component masses (Sec.~\ref{sec:source-frame mass models}), component spins (Sec.~\ref{sec:spin model}), and redshift (Sec.~\ref{sec:redshift model}).


\subsection{Source-frame mass models}
\label{sec:source-frame mass models}

As shown in Eq.~\ref{eq:joint mass}, we separate the source-mass distribution into two one-dimensional mass distributions and a pairing function.
While we change $p_{1D}$ to add additional features, we always consider a simple model for the pairing function introduced in \cite{farah2022bridging}. 
\begin{equation}
    f(m_{1s}, m_{2s}; \beta_1, \beta_2) = \left\{\begin{matrix} (m_{2s}/m_{1s})^{\beta_1} & \text{if } m_{2s} \leq 5\,\rm M_\odot \\ (m_{2s}/m_{1s})^{\beta_2} & \text{if } m_{2s} > 5\,\rm M_\odot \end{matrix}\right.
\end{equation}
This pairing function introduces a discontinuity in $p(m_{1s}, m_{2s}|\Lambda)$ at the fixed switch-point $m_{2s} = 5\,\rm M_\odot$.
It is not clear whether such a discontinuity is appropriate, but it is not thought to affect our conclusions.

Within this framework, we also construct more complicated $p_{1D}$ by starting with a base model (\PDB)
\begin{equation}\label{eq:PDB}
    p_\textsc{PDB}(m|\Lambda) \propto p_\textsc{BPL}(m|\alpha_1, \alpha_2, m_\mathrm{brk}) \, h(m | m_\mathrm{min}, \eta_\mathrm{min}) \, n(m | \gammalowone, \gammahighone, \etalowone, \etahighone, A_1) \, \ell(m | \mmax, \eta_\mathrm{max})
\end{equation}
This combines a broken power-law
\begin{equation}
    p_\textsc{BPL}(m|\alpha_1, \alpha_2, m_\mathrm{brk}) = \left\{\begin{matrix}
        \left(m / m_\mathrm{brk}\right)^{\alpha_1} & \text{if } m < m_\mathrm{brk} \\
        \left(m / m_\mathrm{brk}\right)^{\alpha_2} & \text{if } m \geq m_\mathrm{brk}
    \end{matrix}\right.
\end{equation}
with Butterworth low-pass and high-pass filters
\begin{align}
    \ell(m|k, \eta) & = (1 + (m / k)^\eta )^{-1} \\
    h(m|k, \eta) & = 1 - \ell(m|k, \eta) \\
\end{align}
and a notch filter
\begin{equation}
    n(m|\gammalowone, \gammahighone, \etalowone, \etahighone, A_1) = 1 - A_1 \, h(m|\gammalowone, \etalowone) \, \ell(m|\gammahighone, \etahighone)
\end{equation}
Depending on the sign of $A_1$, the notch filter can either remove a notch ($A_1 > 0$) or add a peak ($A_1 < 0$).

We then extend this model to include multiple peaks and dips.
\PDBP~adds a single additional peak through a multiplicative filter.
\begin{equation}\label{eq:PDBP}
    p_{\PDB\times\textsc{P}}(m|\Lambda) = p_\PDB(m|\Lambda) \times (1 + c_2 \mathcal{G}(m|\peaktwo,\sigtwo))
\end{equation}
where $\mathcal{G}(m|\mu,\sigma)$ is a Gaussian distribution with mean $\mu$ and standard deviation $\sigma$.
\begin{equation}
    \mathcal{G}(m|\mu,\sigma) = \frac{1}{\sqrt{2\pi\sigma^2}} \exp\left(-\frac{(m-\mu)^2}{2\sigma^2}\right)
\end{equation}

Alternatively, \PDBpP~(Sec.~\ref{sec:pdb_p}) constructs a mixture model.
\begin{equation}\label{eq:PDBpP}
    p_{\PDB+\textsc{P}}(m|\Lambda) = (1 - \lambda_2) p_\PDB(m|\Lambda) + \lambda_2 \mathcal{G}(m|\peaktwo,\sigtwo)
\end{equation}

\DoubleDip~extends \PDB~by adding another notch filter at high masses.
\begin{equation}
    p_\DoubleDip(m|\Lambda) = p_\PDB(m|\Lambda) \times n(m|\gammalowtwo, \gammahightwo, \etalowtwo, \etahightwo, A_2)
\end{equation}
while \MultiPDB~instead extends \PDB~by adding a multiplicative filter containing two Gaussian peaks.
\begin{equation}\label{eq:1d-mass}
    p_\MultiPDB(m|\Lambda) = p_\PDB(m|\Lambda) \times (1 + c_1 \mathcal{G}(m|\peakone, \sigone) +  c_2 \mathcal{G}(m|\peaktwo, \sigtwo) )
\end{equation}


\subsection{Spin Model}
\label{sec:spin model}

We assume both components' spins are independently and identically distributed uniformly in magnitude and isotropically in orientation.
The distribution over Cartesian spin components is therefore
\begin{equation}
    p(\vec{s}_1, \vec{s}_2|\Lambda) \propto \frac{1}{|\vec{s}_1|^2 |\vec{s}_2|^2}
\end{equation}


\subsection{Redshift Model}
\label{sec:redshift model}

As discussed in Sec.~\ref{sec:mass distributions}, we assume the merger rate follows the star formation rate (Eq.~\ref{eq:p of z}).
We follow \citet{Fishbach_2018} and define
\begin{equation}
    \Phi(z) = 0.015 \left( \frac{(1+z)^{2.7}}{1+[(1+z) / 2.9]^{5.6}} \right) \, \frac{M_{\odot}}{\mathrm{yr} \, \mathrm{Mpc}^{3}}
\end{equation} 
The rest of our redshift model implicitly depends on the flat $\Lambda$CDM cosmology assumed through the comoving volume ($V_c$).
See Table~\ref{tab:prior} for the precise values of our cosmological model.


\section{Priors and Posteriors}
\label{sec:priors}

Tables \ref{tab:prior} and~\ref{tab:other priors} list the priors assumed within our analysis.
All priors are uniform over a restricted range, and we denote the uniform distribution between $X$ and $Y$ as U($X$, $Y$).

Table~\ref{tab:posteriors} lists posterior medians and 90\% symmetric credible regions for all model parameters for all models.
It additionally lists posterior credible regions for the model-independent summary statistics introduced in Sec.~\ref{sec:nonparametric features}.

\begin{table*}
    \begin{center}
    \caption{
        Hyper-priors for parameters that are common to each mass model.
        We denote a uniform distribution between $X$ and $Y$ as $U(X, Y)$.
        For fixed parameters, we simply report the value assumed.
        }
    \label{tab:prior}
    \setlength{\tabcolsep}{2pt} 
    \renewcommand{\arraystretch}{1.2} 
    \begin{tabular}{c c l p{4.5cm} c}
        \toprule
         & \multicolumn{2}{c}{Parameter} & Description & Prior \\
        
        \hline

        \multirow{5}{*}{Cosmology}
            & $H_0$ & [$\mathrm{km\:s}^{-1}\mathrm{Mpc}^{-1}$] & Present expansion rate & U(40,120) \\
        \cline{2-5}
            & $\Omega_m$ & & Matter density & U(0,1) \\
        \cline{2-5}
            & $\Omega_r$ & & Radiation density & 0.001 \\
        \cline{2-5}
            & $\Omega_k$ & & Curvature & 0 \\
        \cline{2-5}
            & $\Omega_\Lambda$ & & Cosmological constant & $1 - \Omega_m - \Omega_r - \Omega_k$ \\
        \hline

        \multirow{2}{*}{Pairing Function}
            & $\beta_1$ & & Spectral index below $5 \rm M_\odot$ & U(0,10) \\
        \cline{2-5} 
            & $\beta_2$ & & Spectral index above $5 \rm M_\odot$ & U(0,10) \\
            
        \hline

        \multirow{3}{*}{Broken Power-Law}
            & $\alpha_1$ & & Spectral index below $m_{\mathrm{brk}}$ & U(-5,5) \\
        \cline{2-5} 
            & $\alpha_2$ & & Spectral index above $m_{\mathrm{brk}}$ & U(-5,5) \\
        \cline{2-5} 
            & $m_{\mathrm{brk}}$ & $[\rm M_\odot]$ & Dividing point for $\alpha_1$ and $\alpha_2$ & U(2,5) \\
        
        \hline

        \multirow{2}{*}{Highpass Filter}
            & $m_{\mathrm{min}}$ & $[\rm M_\odot]$ & Roll-off scale for low masses & U(0.5,1.2) \\
        \cline{2-5} 
            & $\eta_{\mathrm{min}}$ & & Sharpness of the roll-off at $m_{\mathrm{min}}$ & U(25,50) \\
        
        \hline

        \multirow{2}{*}{Lowpass Filter}
            & \mmax & $[\rm M_\odot]$ & Roll-off scale for high masses & U(35,100) \\
        \cline{2-5} 
            & $\eta_{\mathrm{max}}$ & & Sharpness of the roll-off at \mmax & U(0,10) \\
        
        \hline

        \multirow{5}{*}{Low-Mass Notch}
            & \gammalowone & $[\rm M_\odot]$ & Lower edge of low-mass notch & U(2.3,4) \\
        \cline{2-5} 
            & \etalowone & & Sharpness of the roll-off at \gammalowone & U(0,50) \\
        \cline{2-5} 
            & \gammahighone & $[\rm M_\odot]$ & Upper edge of low-mass notch & U(4,8) \\
        \cline{2-5} 
            & \etahighone & & Sharpness of the roll-off at \gammahighone & U(0,50) \\
        \cline{2-5} 
            & $A_1$ & & Depth of low-mass notch & U(0,1) \\

        \bottomrule
    \end{tabular}
    \end{center}
\end{table*}

\begin{table*}
    \begin{center}
    \caption{
        Additional hyper-priors for each mass model.
        We denote a uniform distribution between $X$ and $Y$ as $U(X, Y)$.
        }
    \label{tab:other priors}
    \setlength{\tabcolsep}{2pt} 
    \renewcommand{\arraystretch}{1.2} 
    \begin{tabular}{c c l p{4.75cm} c c c c c }
        \toprule
         & \multicolumn{2}{c}{Parameter} & Description & \PDB & \PDBpP & \PDBP & \DoubleDip & \MultiPDB \\
        
        \hline

        \multirow{5}{*}{High-Mass Notch}
            & \gammalowtwo & $[\rm M_\odot]$ & Lower edge of high-mass notch
            & - & - & - & U(6,60) & - \\
        \cline{2-9} 
            & \etalowtwo & & Sharpness of the roll-off at \gammalowtwo
            & - & - & - & U(0,50) & - \\
        \cline{2-9} 
            & \gammahightwo & $[\rm M_\odot]$ & Upper edge of high-mass notch
            & - & - & - & U(6,60) & - \\
        \cline{2-9} 
            & \etahightwo & & Sharpness of the roll-off at \gammahightwo
            & - & - & - & U(0,50) & - \\
        \cline{2-9} 
            & $A_2$ & & Depth of high-mass notch
            & 0 & 0 & 0 & U(0,1) & 0 \\

        \hline

        \multirow{3}{*}{Low-Mass Peak}
            & \peakone & $[\rm M_\odot]$ & Location of low-mass peak
            & - & - & - & - & U(6,12) \\
        \cline{2-9} 
            & \sigone & $[\rm M_\odot]$ & Width of low-mass peak
            & - & - & - & - & U(1,40) \\
        \cline{2-9} 
            & $c_1$ & & Height of the low-mass peak
            & 0 & 0 & 0 & 0 & U(0,100) \\
            
        \hline

        \multirow{4}{*}{High-Mass Peak}
            & \peaktwo & $[\rm M_\odot]$ & Location of high-mass peak
            & - & U(20,60) &U(20,60) & - & U(20,60) \\
        \cline{2-9} 
            & \sigtwo & $[\rm M_\odot]$ & Width of high-mass peak
            & - & U(1,40) & U(1,40) & - & U(1,40) \\
        \cline{2-9} 
            & $c_2$ & & Height of high-mass peak
            & 0 & 0 & U(0,100) & 0 & U(0,100) \\
        \cline{2-9} 
            & $\lambda_2$ & & Mixing frac. of high-mass peak
            & 0 & U(0,1) & 0 & 0 & 0 \\

        \bottomrule
    \end{tabular}
    \end{center}
\end{table*}

\begin{table*}
    \centering
    \caption{
        Posterior medians and 90\% symmetric credible regions of all hyper-parameters. We have added the term $\frac{c}{\sigma^{\mathrm{peak}} \sqrt{2 \pi}}$. It should be interpreted as an amplitude representing the number of times the peak is higher than the surrounding mass distribution. 
    }
    \label{tab:posteriors}
    \setlength{\tabcolsep}{4pt} 
    \renewcommand{\arraystretch}{1.2} 
    \begin{tabular}{c cl c c c c}
\toprule
& & Parameter & \PDB & \PDBP & \DoubleDip & \MultiPDB \\
\midrule
\multirow{2}{*}{Cosmology}
& $H_0$ & [$\mathrm{km\:s}^{-1}\mathrm{Mpc}^{-1}$] & \PDBHzero & \PDBPHzero & \DoubleDipHzero & \MultiPDBHzero \\
\cline{2-7}
& $\Omega_{m}$ & & \PDBOmzero & \PDBPOmzero & \DoubleDipOmzero & \MultiPDBOmzero \\
\hline
\multirow{2}{*}{Pairing Function}
& $\beta_1$ & & \PDBbetaone & \PDBPbetaone & \DoubleDipbetaone & \MultiPDBbetaone \\
\cline{2-7}
& $\beta_2$ & & \PDBbetatwo & \PDBPbetatwo & \DoubleDipbetatwo & \MultiPDBbetatwo \\
\hline
\multirow{3}{*}{Broken-Power-Law}
& $\alpha_1$ & & \PDBalphaone & \PDBPalphaone & \DoubleDipalphaone & \MultiPDBalphaone \\
\cline{2-7}
& $\alpha_2$ & & \PDBalphatwo & \PDBPalphatwo & \DoubleDipalphatwo & \MultiPDBalphatwo \\
\cline{2-7}
& $m_{\mathrm{brk}}$&[$M_{\odot}$] & \PDBmbreak & \PDBPmbreak & \DoubleDipmbreak & \MultiPDBmbreak \\
\hline
\multirow{2}{*}{High-Pass Filter}
& $m_{\mathrm{min}}$&[$M_{\odot}$] & \PDBmmin & \PDBPmmin & \DoubleDipmmin & \MultiPDBmmin \\
\cline{2-7}
& $\eta_{\mathrm{min}}$ & & \PDBetamin & \PDBPetamin & \DoubleDipetamin & \MultiPDBetamin \\
\hline
\multirow{2}{*}{Low-Pass Filter}
& $m_{\mathrm{max}}$& [$M_{\odot}$] & \PDBmmax & \PDBPmmax & \DoubleDipmmax & \MultiPDBmmax \\
\cline{2-7}
& $\eta_{\mathrm{max}}$&  & \PDBetamax & \PDBPetamax & \DoubleDipetamax & \MultiPDBetamax \\
\hline
\multirow{5}{*}{Low-Mass Notch}
& $\gamma^{\mathrm{low}}_1$& [$M_{\odot}$] & \PDBgammalow & \PDBPgammalow & \DoubleDipgammalow & \MultiPDBgammalow \\
\cline{2-7}
& $\eta^{\mathrm{low}}_1$ & & \PDBetalow & \PDBPetalow & \DoubleDipetalow & \MultiPDBetalow \\
\cline{2-7}
& $\gamma^{\mathrm{high}}_1$& [$M_{\odot}$] & \PDBgammahigh & \PDBPgammahigh & \DoubleDipgammahigh & \MultiPDBgammahigh \\
\cline{2-7}
& $\eta^{\mathrm{high}}_1$ & & \PDBetahigh & \PDBPetahigh & \DoubleDipetahigh & \MultiPDBetahigh \\
\cline{2-7}
& $A_1$ & & \PDBA & \PDBPA & \DoubleDipA & \MultiPDBA \\
\hline
\multirow{5}{*}{High-Mass Notch}
& $\gamma^{\mathrm{low}}_2$& [$M_{\odot}$] & - & - & \DoubleDipgammalowtwo & - \\
\cline{2-7}
& $\eta^{\mathrm{low}}_2$ & & - & - & \DoubleDipetalowtwo & - \\
\cline{2-7}
& $\gamma^{\mathrm{high}}_2$& [$M_{\odot}$] & - & - & \DoubleDipgammahightwo & - \\
\cline{2-7}
& $\eta^{\mathrm{high}}_2$ & & - & - & \DoubleDipetahightwo & - \\
\cline{2-7}
& $A_2$ & & - & - & \DoubleDipAtwo & - \\
\hline
\multirow{3}{*}{Low-Mass Peak}
& $\mu^{\mathrm{peak}}_1$& [$M_{\odot}$] & - & - & - & \MultiPDBmupeakone \\
\cline{2-7}
& $\sigma^{\mathrm{peak}}_1$& [$M_{\odot}$] & - & - & - & \MultiPDBsigpeakone \\
\cline{2-7}
& $c_1$ & & - & - & - & \MultiPDBpeakconstantone \\
\cline{2-7}
& $\dfrac{c_1}{\sigma^{\mathrm{peak}}_1 \sqrt{2 \pi}}$ & & - & - & - & \MultiPDBpeakconstantoneNORMED \\
\hline
\multirow{3}{*}{High-Mass Peak}
& $\mu^{\mathrm{peak}}_2$& [$M_{\odot}$] & - & \PDBPmupeak & - & \MultiPDBmupeaktwo \\
\cline{2-7}
& $\sigma^{\mathrm{peak}}_2$& [$M_{\odot}$] & - & \PDBPsigpeak & - & \MultiPDBsigpeaktwo \\
\cline{2-7}
& $c_2$ & & - & \PDBPpeakconstant & - & \MultiPDBpeakconstanttwo \\
\cline{2-7}
& $\dfrac{c_2}{\sigma^{\mathrm{peak}}_2 \sqrt{2 \pi}}$ & & - & \PDBPpeakconstantNORMED & - & \MultiPDBpeakconstanttwoNORMED \\
\hline
\multirow{5}{*}{Summary Statistics}
& $\hat{\mu}^{7:11}_1$& [$M_{\odot}$] & \PDBmuhatvalsonezero & \PDBPmuhatvalsonezero & \DoubleDipmuhatvalsonezero & \MultiPDBmuhatvalsonezero \\
\cline{2-7}
& $\hat{\sigma}^{7:11}_1$& [$M_{\odot}$] & \PDBsigmahatvalsonezero & \PDBPsigmahatvalsonezero & \DoubleDipsigmahatvalsonezero & \MultiPDBsigmahatvalsonezero \\
\cline{2-7}
& $\hat{\mu}^{25:40}_2$& [$M_{\odot}$] & \PDBmuhatvals & \PDBPmuhatvals & \DoubleDipmuhatvals & \MultiPDBmuhatvals \\
\cline{2-7}
& $\hat{\sigma}^{25:40}_2$& [$M_{\odot}$] & \PDBsigmahatvals & \PDBPsigmahatvals & \DoubleDipsigmahatvals & \MultiPDBsigmahatvals \\
\cline{2-7}
& $m_{99}$& [$M_{\odot}$] & \PDBmninenine & \PDBPmninenine & \DoubleDipmninenine & \MultiPDBmninenine \\
\bottomrule
    \end{tabular}
\end{table*}

\clearpage
\bibliography{refs}
\bibliographystyle{aasjournal}

\end{document}